\documentclass[11pt]{article}
\pdfoutput=1
\usepackage{amsmath,amsfonts,amssymb,mathtools}
\usepackage{graphicx}
\usepackage[dvipsnames,svgnames,table]{xcolor}
\usepackage{booktabs}
\usepackage{makecell}
\usepackage{adjustbox}
\PassOptionsToPackage{sort&compress}{natbib}
\usepackage{jheppub}
\hypersetup{colorlinks=true,linkcolor=BrickRed,citecolor=BrickRed,urlcolor=BrickRed,linktoc=page}
\newcommand{\MP}{M_{P}}
\newcommand{\Tmax}{T_{\max}}
\newcommand{\EN}{E_{N}}
\newcommand{\QN}{Q_{N}}
\newcommand{\dd}{\mathrm{d}}
\newcommand{\Tc}{T_{c}}
\newcommand{\lS}{\ell_{\rm S}}
\newcommand{\lc}{\ell}
\newcommand{\arXivold}[2]{\href{http://arxiv.org/abs/#1}{{\tt arXiv:#1 [#2]}}}

\title{Thermal Festina Lente}

\author[a]{Fayez Abu-Ajamieh,}
\author[b]{Nobuchika Okada,}
\author[c]{Roman Pasechnik,}
\author[d]{Victoria Venken,}
\author[e]{and Zhi-Wei Wang}

\affiliation[a]{Formerly: The Indian Institute of Science, Bangalore, India}
\affiliation[b]{Department of Physics and Astronomy, University of Alabama, Tuscaloosa, AL 35487, USA}
\affiliation[c]{Department of Physics, Lund University, SE-223 62 Lund, Sweden}
\affiliation[d]{Kavli Institute for the Physics and Mathematics of the Universe (WPI), The University of Tokyo Institutes for Advanced Study, The University of Tokyo, Kashiwa, Chiba 277-8583, Japan}
\affiliation[e]{School of Physics, The University of Electronic Science and Technology of China, Chengdu 611731, China}

\emailAdd{fayezabuajamieh@gmail.com}
\emailAdd{okadan@ua.edu}
\emailAdd{roman.pasechnik@fysik.lu.se}
\emailAdd{victoria.venken@ipmu.jp}
\emailAdd{zhiwei.wang@uestc.edu.cn}

\abstract{

According to the Festina Lente bound, particles charged under a $U(1)$ gauge field cannot be too light. Otherwise, the largest charged black holes of de Sitter space would discharge in a way that leads to a crunching singularity in violation of cosmic censorship. The Festina Lente bound has previously been extended to also include a background of thermal radiation, but where vacuum energy still dominates cosmic evolution, which imposes a maximal temperature $T\leq\Tmax=(cV/\sigma)^{1/4}$. 
We extend this analysis by accounting for thermal enhancement of the discharge process and making the analysis more quantitative. The result is a bound $m\gtrsim\lc (cV/\sigma)^{1/4}$ on the masses of charged particles. Here, $\lc$ is a term that depends logarithmically on the gauge coupling and particle charge. At weak coupling this leads to a significantly stronger constraint than the $\sqrt{gq}$ scaling of the original Festina Lente bound.
 
As a result, the present vacuum energy requires every particle of charge $10^{-26}\lesssim gq\lesssim e$ to be heavier than $0.03$--$0.2$\,eV, provided its gauge boson is lighter than the Hubble scale. An unbroken gauged $B-L$ with Dirac neutrinos gives $m_{\nu,{\rm lightest}}\gtrsim10$--$43$\,meV for $10^{-29}\lesssim g_{B-L}\lesssim10^{-24}$, in tension with the DESI\,DR2 limit once $g_{B-L}\gtrsim7\times10^{-28}$. The same limit bounds a neutrino millicharge to $\epsilon_\nu\lesssim2\times10^{-27}$. We also revisit the Standard Model effective field theory constraints from the electroweak phase transition. During slow-roll inflation the Standard Model electron violates the bound over the whole range of Hubble rates allowed by the CMB tensor limit and by nucleosynthesis, unless the photon is Higgsed, its coupling is suppressed, or one considers very fine-tuned versions of Higgs inflation.

}

\begin{document}
\maketitle

\section{Introduction}\label{intro}

The Festina Lente (FL) bound~\cite{Montero:2019ekk,Montero:2021otb} states that in de Sitter space with vacuum energy $V$, all particles of charge $q$ under a $U(1)$ gauge field with coupling $g$ should have a mass $m$ such that
\begin{equation}\label{eq:FLoriginal}
    m^4 \gtrsim 2 (g q)^2 V\,,\qquad\text{equivalently}\qquad m^2\gtrsim\sqrt6\,gq\,\MP H\,,
\end{equation}
where $\MP=2.4\times10^{18}$\,GeV is the reduced Planck mass and $H=\sqrt{V/3}/\MP$ the Hubble rate of the de Sitter background~\cite{Montero:2021otb}. This means that a charged particle that is too light would let the largest black holes of de Sitter space discharge almost at once, and what would be left behind is a singular geometry rather than a black hole. Requiring that this does not happen puts a floor, set by the vacuum energy, under the mass of every charged particle.

\begin{figure}[t]
    \centering

    \includegraphics[width=0.5\linewidth]{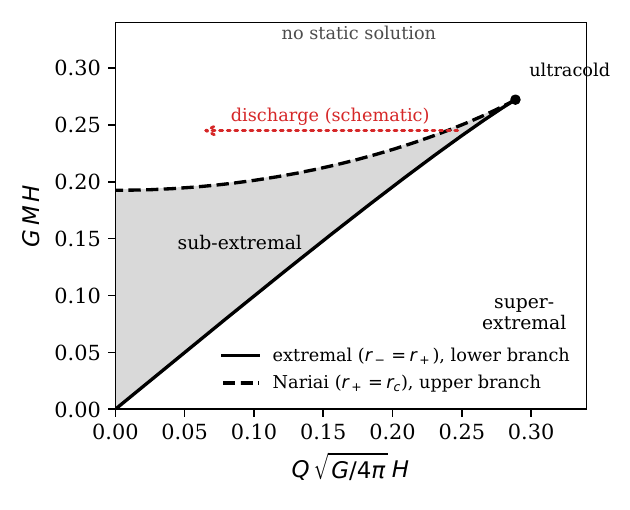}
    \caption{The mass $M$ versus charge $Q$ of black holes in de Sitter space. The black holes inside the sharkfin shape are subextremal. Below the lower boundary the solutions are superextremal, and above the upper boundary no static solution exists. The lower boundary of the sharkfin (solid) consists of the usual extremally charged black holes, while the upper boundary (dashed) consists of charged Nariai black holes where the black hole horizon coincides with the cosmic horizon. The two branches meet at the ultracold point, where the electric field of the Nariai branch is maximal, $E_N=\sqrt6\MP H$ (along the extremal branch the field keeps growing toward smaller black holes). The curve is the exact locus of degenerate horizons of the Reissner--Nordstr\"om--de Sitter solution, $(Q\sqrt{G/4\pi}\,H,\,GMH)=(r\sqrt{1-3r^2},\,r-2r^3)$ with $r=r_{\rm hor}H$ and $Q$ the charge in canonical (Heaviside--Lorentz) units (cf.~\cite{Montero:2019ekk}). The dotted arrow is a schematic indication of the rapid-discharge process, which leads out of the region of static solutions and is not a quasistatic motion in this diagram.}
    \label{fig:BHdiagram}
\end{figure}

The central argument for this bound is based on the cosmic censorship conjecture~\cite{Penrose:1969pc}, which forbids the formation of singularities that are not hidden behind a horizon. Because de Sitter space has a cosmic horizon, black holes have a maximum size of order the Hubble scale, beyond which no static black-hole solution exists. These largest possible black holes are known as charged Nariai black holes~\cite{Nariai:1950a,Nariai:1950b} (see~\cite{Romans:1991nq,Bousso:1996au} for the charged solutions and their nucleation in de Sitter space). The diagram of charged black hole solutions in de Sitter space is shown in Figure~\ref{fig:BHdiagram}. The subextremal black holes form a sharkfin shape, where charged black holes can have a larger mass than neutral black holes before reaching the Nariai limit.

This provides a potential pathway to produce a singularity as depicted by the dotted arrow in Figure~\ref{fig:BHdiagram}: one may imagine starting with a charged near-Nariai black hole of cosmic size. If the black hole can discharge very rapidly without losing sufficient mass at the same time, one may end up in a singular regime.

Let us stress that this is only a cartoon picture of the actual physics: each point in Figure~\ref{fig:BHdiagram} corresponds to a static classical black hole solution. To think of the evaporation of the black hole as a flow in this diagram, the evaporation must happen quasistatically, which no longer holds for rapid discharge. We refer to~\cite{Montero:2019ekk} for the detailed analysis of the evaporation process\footnote{The main physical difference from the quasistatic picture is that, according to~\cite{Montero:2019ekk}, the discharge converts the electric field into a cloud of neutral radiation in a time much shorter than the Hubble time. The backreaction of this radiation leads to a metric different from the black hole in empty space. The energy density in the radiation produces a big-crunch-type singularity whose formation is guaranteed by singularity theorems. Whether this crunch should be regarded as a violation of cosmic censorship has recently been questioned in~\cite{Aalsma:2023mkz}. We return to this point in Section~\ref{sec:Discussion}. In Section~\ref{sec:criterion} and Appendix~\ref{app:discharge} we show that the field energy need not end up in radiation for this to happen. What starts the contraction is the loss of the tension of the field, whatever the field turns into. We integrate three cases, the accelerated charges themselves, radiation and dust, and in all three the throat reaches a curvature singularity in a time of order $H^{-1}$. The time scale, unlike the onset, does depend on what the energy becomes: a component whose pressure is only marginally above $-\rho$ contracts the throat arbitrarily slowly. Whether the singularity is naked, as argued in~\cite{Montero:2019ekk}, is a question about the inhomogeneous evolution, which these models do not settle.}. At a qualitative level the argument runs as follows. For a constant electric field $E$ in flat space, the Schwinger process~\cite{Schwinger:1951nm} has a pair-production rate exponentially suppressed as $\exp(-\pi m^2/gqE)$. This exponential suppression is lost when the particle becomes too light. Requiring the discharge to be exponentially suppressed implies $m^2 \gtrsim g q E$. Using the electric field $E$ for a black hole near the top right corner in Figure~\ref{fig:BHdiagram}, $E_N=\sqrt6\,\MP H$ at the ultracold point~\cite{Montero:2019ekk}, and accounting for the physics in de Sitter space then leads to eq.~\eqref{eq:FLoriginal}.

A realistic cosmology has radiation and matter-driven components on top of the vacuum energy driving pure de Sitter evolution. In~\cite{Venken:2023hfa} the Festina Lente argument was extended to account for the presence of a background of homogeneous radiation at a finite temperature $T$. Since the energy density in radiation is $\rho_R = \sigma T^4$ and the vacuum energy is $\rho_V = V$, it was argued that, so long as the vacuum energy dominates,
\begin{equation}
\label{eq:Tbound}
    \frac{\rho_R}{\rho_V} \leq c \implies T \leq \Tmax\equiv\left(\frac{c V}{\sigma}\right)^{1/4}\,,
\end{equation}
where $\sigma = \frac{\pi^{2}}{30}g_{*}(T)$, $g_{*}(T)$ is the number of effectively massless degrees of freedom and $c$ is an $\mathcal{O}(1)$ constant. When this holds, the original Festina Lente argument remains valid, and in~\cite{Venken:2023hfa} it was used to derive bounds on the electroweak and QCD phase transitions and on inflation. Throughout we keep $c$ and $g_*$ explicit in the formulae. For numerical estimates we take $c=1$ and, as the minimal bath compatible with the argument, the two polarizations of the $U(1)$ gauge boson itself, $g_*=2$. For other choices $\Tmax$ scales as $(c/g_*)^{1/4}$, which moves the bound itself, and with it the couplings that a given measurement excludes. Unless stated otherwise, the present-day bounds quoted below and in Sections~\ref{sec:neutrinos} and~\ref{sec:mCP} refer to this benchmark, and those sections also give their dependence on $c$.

What ref.~\cite{Venken:2023hfa} did not account for is that the thermal bath itself changes the way a Nariai black hole discharges: thermal fluctuations assist Schwinger pair creation~\cite{Brown:2015kgj,Gould:2017fve,Gould:2018ovk}, and, more importantly, the bath contains a population of the charged particles themselves, which are accelerated by the black-hole field and neutralize it. Working out both effects and their phenomenological implications is the goal of this paper.

In Section~\ref{sec:finiteT} we formulate the finite-temperature version of the argument. The key step is to replace the qualitative requirement that ``the Schwinger rate must be exponentially suppressed'' by the quantitative criterion that a near-ultracold charged Nariai black hole must not lose an order-one fraction of its charge $\QN\sim\MP/(gqH)$ within a Hubble time. For a thermal population of charged particles, losing that charge simply means that the carriers between the two horizons screen the field. In the homogeneous models of the throat that we solve, the energy of the field then goes into those carriers and the throat contracts to a curvature singularity in a time of order $H^{-1}$. We take this criterion, restricted to near-ultracold black holes, as our definition of rapid discharge. Applied at zero temperature this reproduces eq.~\eqref{eq:FLoriginal} up to a logarithm. Applied to an equilibrium bath of temperature $T$ it gives the \emph{thermal Festina Lente bound}
\begin{equation}\label{eq:TFL_intro}
    m\;\gtrsim\;\lc\,T\,,\qquad \lc=\ln\!\left[\frac{2g_s\,gq\,(mT/2\pi)^{3/2}}{\sqrt6\,\MP H^2}\right],
\end{equation}
which follows from the requirement that the Boltzmann-suppressed population $n\propto e^{-m/T}$ of the charged species be too dilute to conduct the black-hole charge away\footnote{One caveat, as we shall see, is that the derivation of this bound requires $gq$ larger than $(gq)_{\min}=\frac{2\sqrt6\,\pi^2}{3\zeta(3)g_s}\big(\frac{\sigma}{3c}\big)^{3/4}\sqrt{H/\MP}$. At weaker coupling the discharge channel does not operate.}. Here, $g_s$ is the number of spin states. For the maximal bath $T=\Tmax$ the bound is of order $V^{1/4}$, and the coupling enters only through the logarithm: at the present value of the vacuum energy $\lc$ ranges from $74$ for the electron charge to $17$ for $gq=10^{-24}$, and the bound disappears only for $gq\lesssim10^{-30}$. Thermally assisted pair creation through photon fusion gives a parametrically similar but weaker condition wherever the two calculations can both be trusted. The conduction bound does not rest on that rate at all, and we make no claim about channels we have not computed. By contrast, the naive criterion ``exponent $\gtrsim1$'' would give $m\gtrsim T/2$, and the criterion ``exponent $\gtrsim\pi$'' that reproduces eq.~\eqref{eq:FLoriginal} at $T\to0$ would give $m\gtrsim\pi T/2$. We show that both miss the dominant physics and the potentially large logarithm. The order-one choices that enter the criterion shift the logarithms by additive constants, which we estimate to be $\pm3$ in total. We carry that as a sensitivity band on the normalization of the current.

A remarkable consequence is that if $B-L$ is an unbroken gauge symmetry whose gauge boson is lighter than the Hubble scale, the lightest neutrino must be heavier than $10$--$43$\,meV for any $10^{-29}\lesssim g_{B-L}\lesssim10^{-24}$, which turns the numerical coincidence $m_\nu\sim V^{1/4}$ noted in~\cite{Montero:2019ekk,Montero:2021otb} into a testable statement.

Section~\ref{sec:particlephysics} works out the particle-physics implications. We first discuss neutrinos and gauged $B-L$, then millicharged particles, for which FL-type constraints were analyzed earlier in~\cite{Ban:2022jgm,Montero:2022jrc,Abu-Ajamieh:2025sul}. Our bound remains nontrivial down to charges many orders of magnitude below the reach of any laboratory test. Finally we revisit the constraints on the dimension-six operators of the Standard Model effective field theory (SMEFT) that follow from requiring the electroweak phase transition to take place in a radiation-dominated universe~\cite{Venken:2023hfa,Lee:2021cor}, using the physical number of relativistic degrees of freedom, and find that they do not go beyond the absence of a symmetric de Sitter minimum of the Higgs potential.

Section~\ref{sec:inflation} deals with the early universe. If the early universe had a period of slow-roll inflation, then the vacuum energy would have been very large and our conjecture thus leads to a severe constraint\footnote{Zero-temperature FL already strongly constrains inflation as was stressed in the original paper \cite{Montero:2019ekk}.}. The Standard Model electron violates it for every inflationary Hubble rate between $5\times10^{-27}$\,GeV and $2.7\times10^{15}$\,GeV, and that interval contains the whole range allowed by the CMB limit on tensor modes and by nucleosynthesis. 

If our conjecture is correct, this leaves two possible options: Either slow-roll inflation does not provide a valid description of the early universe or we must heavily modify the Standard Model sector during inflation. A natural option would be to use the Higgs as the inflaton to ensure all charged matter fields are sufficiently heavy. However, we shall see that vanilla Higgs inflation does not do the job and a more fine-tuned potential is needed. Another option is that the inflaton couples to the gauge fields such that during inflation all gauge fields were Higgsed\footnote{A particularly compelling proposal along these lines within the MSSM appeared in \cite{Nakamura:2026qvp} while we were finalizing this manuscript.}, confined, or so weakly coupled that our bound is satisfied.

We conclude in Section~\ref{sec:Discussion} with a discussion and outlook for further work. Technical details are collected in the appendices: the worldline-instanton derivation of the thermal Schwinger exponent and its high-temperature expansion (Appendix~\ref{app:instanton}), the discharge-time estimates (Appendix~\ref{app:discharge}), the tree-level SMEFT formulae (Appendix~\ref{app:smeft}), the reheating bath and the domain of validity of the conjecture (Appendix~\ref{sec:reheatingtop}) and a brief review of Higgs inflation (Appendix~\ref{sec:higgsreview}).
\section[Festina Lente in a thermal bath]{Festina Lente in a thermal bath}
\label{sec:finiteT}
As we saw in the introduction, the main premise of FL is that charged black holes of cosmological size in de Sitter space, known as Nariai black holes, should not discharge too quickly via Schwinger pair production, since a rapid discharge is conjectured to lead to a crunch of the Nariai throat, the region between the two horizons, that is incompatible with cosmic censorship. That the singularity reached is naked is an assumption of that argument, to which we return in Section~\ref{sec:criterion} and in the Discussion. To generalize this argument to finite temperature, consider a Nariai black hole in a thermal bath. Two things change with respect to the vacuum case: pair creation is assisted by the thermal fluctuations of the bath, and the bath contains, in equilibrium, a population of the charged particles themselves, which respond to the electric field of the black hole. Both effects have to be included.

A full analysis at this point would require a calculation of the Schwinger rate in an approximately $dS_2$ geometry with a background of thermal radiation added. This analysis would be further complicated by the fact that as the universe expands, the radiation cools and dilutes, and thus the whole process is time-dependent. We will not deal with these subtleties here and will take the flat-space thermal rates as a good enough approximation to see when the rapid discharge occurs. This is justified by the following:
\begin{itemize}
    \item Deep in the rapid-discharge regime, the discharge time is much shorter than $H^{-1}$, so the geometry and bath change little during the process. At the boundary $t_d\sim H^{-1}$ this separation of scales is lost. Our threshold is therefore a diagnostic based on the local current estimate and the assumptions stated below, rather than a solution of the evolving thermal bath.
    \item So long as the length scale over which a pair is formed is short compared to the Hubble scale, curvature effects will be small and we can think of the pair production process as happening locally in an approximately flat patch. That scale is the radius $R=m/(gq\EN)$ of the worldline instanton rather than the Compton wavelength. The conditions under which the flat-space treatment applies are stated in Section~\ref{sec:pair}. In deriving the original FL bound in~\cite{Montero:2019ekk}, accounting for the $dS_2$ background in calculating the Schwinger pair production was done using the work of~\cite{Frob:2014zka} (see~\cite{Kobayashi:2014zza} for the four-dimensional de Sitter case). In the end, however, it was clear there that the correct parametric result could be gleaned from the flat-space calculation. Thus, and since we are only interested in obtaining parametric results here, this is an acceptable approximation. We will also see that the final bound on the mass of charged particles is generically parametrically larger than the Hubble scale, which justifies the approximation further.
\end{itemize}

\subsection{Setup and the discharge criterion}\label{sec:criterion}

We consider an effective field theory coupled to gravity with a positive vacuum energy $V=3\MP^2H^2$ and a $U(1)$ gauge field with coupling $g$, whose gauge boson is lighter than $H$, under which a particle of mass $m$, integer charge $q$ and $g_s$ spin states is charged. The reference field strength is the largest electric field available on the Nariai branch of Figure~\ref{fig:BHdiagram}, that of the ultracold Nariai black hole~\cite{Montero:2019ekk}, so called because its horizons degenerate and its Hawking temperature vanishes, which has nothing to do with the temperature $T$ of the bath,
\begin{equation}\label{eq:EN}
    \EN=\sqrt6\,\MP H=\sqrt{2V}\,,\qquad r_N=\frac{1}{\sqrt6\,H}\,,
\end{equation}
where $r_N$ is the common radius of the horizons (Appendix~\ref{app:discharge}). In units of the elementary charge $gq$ this black hole carries
\begin{equation}\label{eq:QN}
    \QN=\frac{4\pi r_N^2\EN}{gq}=\frac{4\pi}{\sqrt6}\,\frac{\MP}{gq\,H}\simeq\frac{1.0\times10^{61}}{gq}\qquad\text{(for the present vacuum energy)}\,,
\end{equation}
where we used $V_0^{1/4}=2.24$\,meV, i.e.\ $H_\Lambda=\sqrt{V_0/3}/\MP=1.2\times10^{-33}$\,eV~\cite{Planck:2018vyg}. Eq.~\eqref{eq:QN} is the reason why the prefactor of any discharge rate matters: for the black hole to discharge ``rapidly'' in the sense of~\cite{Montero:2019ekk}, an order-one fraction of $\QN$ must be removed before the geometry can react, i.e.\ within a Hubble time. The region between the two horizons of a near-Nariai black hole, the throat, is the homogeneous space $dS_2\times S^2$~\cite{Romans:1991nq,Bousso:1996au} (see also~\cite{Montero:2020rpl}), with the electric field along the $dS_2$ direction (Appendix~\ref{app:discharge}). Any current $J$ that develops in it is homogeneous as well. Amp\`ere's law reduces to $\dd(b^2E)/\dd t=-b^2J$ for a sphere of radius $b$, i.e.\ $\dot E=-J$ as long as the sphere is static, and the charge of the black hole, $4\pi b^2E$, is removed in the discharge time $t_d=\EN/J$, where $J$ is the saturated current of the considered process. Since the geometry reacts to the conversion, $t_d$ is a reference time evaluated on the static throat, and the flux history of the self-consistent evolution is given in Appendix~\ref{app:discharge}. What the criterion below refers to is the removal of the charge, not the value of the local field\footnote{Strictly, the time needed to remove a fraction $f$ of the field is $t_f=\int\dd E/J(E)$. For the conduction current of Section~\ref{sec:conduction}, $J$ does not depend on $E$ once the carriers are relativistic, so that $t_f=f\,t_d$. For pair creation the rate depends exponentially on $E$, but the current is carried by the pairs accumulated since the onset of the discharge, and $t_f$ exceeds $t_d$ only by a factor of order $\sqrt S$ (Section~\ref{sec:pair}).}. We therefore adopt the following criterion, which we use both at zero and at finite temperature:
\begin{equation}\label{eq:criterion}
    \text{rapid discharge}\quad\Longleftrightarrow\quad t_d=\frac{\EN}{J}<H^{-1}\,.
\end{equation}
The FL conjecture is the statement that rapid discharge must not be possible for any charged Nariai black hole. We restrict the statement to the near-ultracold ones, $E\simeq\EN$, for the reason given below. Its thermal extension, which we propose here, is:

\medskip
\noindent\emph{Thermal Festina Lente conjecture.} Consider an EFT coupled to gravity with a positive cosmological constant $V$, and any of its equilibrium thermal states of temperature $T\leq\Tmax=(cV/\sigma)^{1/4}$, i.e.\ any state in which the background remains vacuum dominated. In none of these states may a near-ultracold charged Nariai black hole, $E\simeq\EN$, satisfy the rapid-discharge criterion~\eqref{eq:criterion}.

\medskip
\noindent Three comments are in order. First, the states in question are hypothetical: the present universe has $T_{\rm CMB}=0.235$\,meV, whereas $\Tmax(V_0)=2.5$\,meV. The logic is the usual one of swampland-type arguments~\cite{Vafa:2005ui,Palti:2019pca,vanBeest:2021lhn}, which constrain the Lagrangian by demanding consistency of every state of the theory, in the same way in which the FL bound itself relies on the existence of Nariai black holes that have never formed. In particular, we do not require that the charged species could thermalize with the bath dynamically (for the tiny couplings considered below it could not). The equilibrium state is used, like the Nariai black hole itself, as a probe of the effective theory. If the charged species belongs to a sector with its own temperature $T_a\leq T$, the bound below applies with $T\to T_a$. The extremal state that defines the bound is the one with $T_a=\Tmax$. The restriction to near-ultracold black holes matters here. A black hole with $E\ll\EN$ would lose its field even faster, since $t_d\propto E$ at fixed current. Its field, however, carries a negligible fraction of the energy of the throat, so removing it has no gravitational consequence beyond the ordinary instability of the Nariai geometry (Appendix~\ref{app:discharge}). The hole that decides the fate of the geometry is the one whose field energy is an order-one fraction of the energy of the throat, and that is the one that sets the bound, as in~\cite{Montero:2019ekk}. Second, the state with $T=\Tmax$ and $c=1$ is only marginally de Sitter, and the radiation redshifts away within one $e$-fold. Inside the Nariai throat the bath is not a small perturbation either. A homogeneous radiation component with $\rho_R=cV$ has a positive pressure along the field lines, and that alone destabilizes the throat: it collapses through its $dS_2$ direction, with no discharge at all, within $0.77/H$ for $c=1$, $1.7/H$ for $c=0.1$ and $3.3/H$ for $c=0.01$ (Appendix~\ref{app:discharge}). For $c\ll1$ the bath is instead a controlled gravitational perturbation, the throat survives for several Hubble times, and the discharge criterion then refers to a background that would otherwise persist. In the absence of a top-down argument fixing $c$, we nevertheless quote the numbers for $c=1$, the formal envelope of the conjecture, in the spirit of~\cite{Venken:2023hfa}, as this is the absolute maximal value where our arguments have any chance of holding. The criterion~\eqref{eq:criterion} compares the discharge time with $H^{-1}$. Replacing $H^{-1}$ by any other order-one multiple of the Hubble time changes the logarithms derived below by an additive constant of order unity, which is covered by the sensitivity band of Section~\ref{sec:TFLbound}. The choice of $c$ is a different matter: it changes $\Tmax$ by $c^{1/4}$ and therefore the bound itself. For $c=0.1$ the logarithm also decreases, by $1.8$ to $1.9$, so that $m_{\min}$ is multiplied by $0.50$--$0.55$ for $gq\geq10^{-24}$, and the couplings that a given experimental limit excludes move with it (Section~\ref{sec:neutrinos}).

Third, the criterion~\eqref{eq:criterion} is adopted as the definition of rapid discharge because of what the conversion of the field energy does to the geometry. The field sustains the Nariai throat through its tension: in the Kantowski--Sachs form of the geometry~\cite{Kantowski:1966te} the radius $b$ of the $S^2$ obeys $2\ddot b/b+\dot b^2/b^2+1/b^2=\Lambda-8\pi G\,p_\parallel$, where $p_\parallel$ is the pressure along the field lines, and the static throat exists because the field has $p_\parallel=-\rho_E$ (Appendix~\ref{app:discharge}). Suppose now that an order-one fraction of $\rho_E$ is converted into anything with a larger pressure along the field lines: the counter-streaming carriers themselves ($p_\parallel=\rho$), the radiation they may emit ($p_\parallel=\rho/3$), or a plasma oscillation, whose time-averaged $p_\parallel$ is positive. The sphere then starts to contract, initially with $\ddot b/b=-4\pi G\,(p_\parallel+\rho_E)$, and for the conversion channels we integrate it reaches zero size in a time of order $H^{-1}$: $0.65/H$ for the streams, $0.75/H$ for radiation and $0.83/H$ for dust, from the numerical integration of Appendix~\ref{app:discharge}. If only half of the field energy is converted into radiation, the crunch proceeds through the collapse of the $dS_2$ direction instead, after $1.1/H$, and after $0.96/H$ for half conversion into streams. The onset of the contraction follows from the focusing equation $\dd(\dot b/a)/\dd t=-(\rho+p_\parallel)\,b/(2\MP^2a)$ for any source obeying the null energy condition, but the time scale does not: a component whose pressure is only marginally above $-\rho$ contracts the throat arbitrarily slowly ($2.4/H$ for $p=-0.99\rho$ and $2.8/H$ for $p=-0.997\rho$), so the times quoted here are those of the channels we integrate, which are the physically relevant ones. These times also refer to the near-ultracold black holes that set the bound. For a Nariai black hole with a field far below $\EN$ the same equations describe the ordinary gravitational instability of the Nariai geometry~\cite{Bousso:1996au}, on a longer time scale. This is the crunch of~\cite{Montero:2019ekk}, and it does not require the carriers to annihilate or to thermalize into radiation: what has to happen within a Hubble time is the conversion of the field energy, which on the static throat goes together with the removal of the charge that eq.~\eqref{eq:criterion} measures. The homogeneous models establish the onset of the contraction and its time scale. They do not by themselves single out the threshold $t_d=H^{-1}$, since they also crunch for slower discharges ($1.2/H$ for $t_d=2H^{-1}$ and $1.8/H$ for $t_d=10H^{-1}$). The threshold is the condition that the conversion be complete before the inhomogeneous modes of the Nariai geometry, which grow on a comparable time scale, can take over and possibly fragment the throat into black holes (Appendix~\ref{app:discharge}). Whether the singularity is naked, as argued in~\cite{Montero:2019ekk}, is therefore not settled by the homogeneous models, and we adopt the criterion as a conjectural diagnostic that shares this assumption with the original argument. The homogeneity of the throat also means that no charge is lost through the horizons: what leaves the patch of one static observer enters the patch of a neighboring one, and the energy of the field is transferred to the carriers, at the rate $JE$ per unit volume, inside the throat. The subsequent fate of the carriers is a separate question, to which we return in Section~\ref{sec:TFLbound} and Appendix~\ref{app:discharge}.

\subsection[Pair creation at zero and finite temperature]{Pair creation at zero and finite temperature}\label{sec:pair}

There exist many attempts at generalizing the Schwinger pair production to finite temperature~\cite{Kim:2007ra, Kim:2008em, Monin:2009aj, Kim:2010qq, King:2012kd, Fukushima:2014sia, Medina:2015qzc, Gould:2017fve, Korwar:2018euc, Brown:2015kgj}. For a weakly coupled charged particle the leading exponent is by now well established: the thermal enhancement arises from the lens-shaped worldline instanton found by Brown~\cite{Brown:2015kgj}, which was shown in~\cite{Gould:2017fve} to give the leading exponent above a critical temperature to all orders in the exchange of virtual photons between the worldline segments (other approaches, e.g.~\cite{Medina:2015qzc}, differ in the observable and in the thermal ensemble considered, and we do not rely on them). The perturbative photon-fusion prefactor we use below was computed in~\cite{Torgrimsson:2019sjn}. Ref.~\cite{Gould:2018ovk} instead computes the worldline-sphaleron rate in a different regime. We use the leading exponent of~\cite{Brown:2015kgj}, re-derived in Appendix~\ref{app:instanton}. According to~\cite{Brown:2015kgj}, the rate of Schwinger pair production at finite temperature is given by
\begin{equation}\label{eq:Scwhinger-T}
    \Gamma_{T>\Tc} \sim \exp{\Bigg[ -\frac{2m^{2}}{gqE}\sin^{-1}{\Big(\frac{\Tc}{T}\Big)}-\frac{m}{T}\sqrt{1-\frac{\Tc^{2}}{T^{2}}}\Bigg]}\;\equiv\;e^{-S_T(m,T)},
\end{equation}
where the critical temperature $\Tc \equiv \frac{gqE}{2m}$ and $gqE$ is the product of the coupling, the charge and the (canonically normalized) electric field. At $T=\Tc$ the exponent equals the zero-temperature Schwinger exponent, $S_{\Tc}=\pi m^2/(gqE)$, and for $T<\Tc$ it keeps that value.

\paragraph{Zero temperature.} The rate per unit volume of Schwinger pair creation of a spin-$\tfrac12$ particle is $\Gamma=(gqE)^2e^{-S}/4\pi^3$ with $S=\pi m^2/(gqE)$~\cite{Schwinger:1951nm,Affleck:1981bma} (for a complex scalar the prefactor is halved). In the region between the two horizons of a near-Nariai black hole the produced charges are accelerated to the speed of light within a time $m/(gq\EN)\ll H^{-1}$ and swept into the horizons after a residence time of order $H^{-1}$ (Appendix~\ref{app:discharge}), so that the steady-state current density is $J_{\rm p}\simeq2gq\,\Gamma/H$. The criterion~\eqref{eq:criterion} then reads
\begin{equation}\label{eq:pair_criterion}
    t_{\rm p}=\frac{2\pi^3H\,e^{S}}{(gq)^3\EN}<H^{-1}
    \quad\Longleftrightarrow\quad
    S<\lS\equiv\ln\!\left[\frac{\sqrt6}{2\pi^3}\,(gq)^3\,\frac{\MP}{H}\right].
\end{equation}
Demanding that rapid discharge be impossible, $S\geq\lS$, gives the zero-temperature FL bound in the form
\begin{equation}\label{eq:FL_log}
    m^2\;\geq\;\frac{\lS}{\pi}\,gq\,\EN=\frac{\lS}{\pi}\,\sqrt6\,gq\,\MP H\,.
\end{equation}
Replacing the logarithm by $\pi$ reproduces eq.~\eqref{eq:FLoriginal}, i.e.\ the form $m^2\gtrsim gq\EN$ of~\cite{Montero:2019ekk,Montero:2021otb}, in which the order-one factor of the criterion was not kept in the final form of the bound (the discharge-time estimate of~\cite{Montero:2019ekk}, Section~4.1.1 there, is the one evaluated here). For the present vacuum energy the logarithm is large, since $\MP/H_\Lambda=2\times10^{60}$. For the electron charge $\lS=132$, and eq.~\eqref{eq:FL_log} is stronger than eq.~\eqref{eq:FLoriginal} by a factor $(\lS/\pi)^{1/2}=6.5$ in the mass ($9.5$\,meV instead of $1.5$\,meV for $gq=e$). Conversely, $\lS$ becomes negative for $gq<(2\pi^3H/\sqrt6\MP)^{1/3}=2.3\times10^{-20}$: for such small couplings the semiclassical pair-creation current cannot discharge the black hole within a Hubble time even for a massless particle, and no bound follows from pair creation. The magnetic bound $q^2g^2\geq\tfrac32(H/\MP)^2$ of~\cite{Montero:2021otb} (see also~\cite{Huang:2006hc}) is a necessary condition for the black hole to exist, not a sufficient one for its rapid discharge.

Two remarks on the validity of this estimate. The bound is extracted at $S=\lS$, which for the electron charge at the present vacuum energy is $132$: well inside the semiclassical regime. A criterion of the form $S\gtrsim1$ would instead extract the bound at $S$ of order one, where the semiclassical expansion is not controlled. This control is lost in any case as $\lS$ approaches zero at the endpoint, since a positive logarithm need not be a large one. The logarithm $\lS$ of this section and the conduction logarithm $\lc$ of Section~\ref{sec:conduction} are different quantities. At $gq=10^{-24}$, for instance, $\lc=17$ while $\lS=-30$: the zero-temperature criterion is empty there, and only the thermal one applies. At the inflationary scales of Section~\ref{sec:slowroll} the logarithm is much smaller: $\lS=5.6$ at $H_I=10^{13}$\,GeV, and $4.0$ at the upper end of the observationally allowed range, $H_*\simeq4.7\times10^{13}$\,GeV. The threshold mass there therefore carries an order-one sensitivity to the normalization of the current, which we quantify in Section~\ref{sec:TFLbound}. The local strong-field estimate itself remains controlled, with $gq\EN/H^2\simeq4\times10^{4}$ and $R(m)H\simeq6\times10^{-3}$ at that scale. The worldline instanton has radius $R=m/(gq\EN)$ and the flat-space treatment requires $RH<1$, i.e.\ $m<\sqrt6\,gq\MP$. At the bound~\eqref{eq:FL_log} this is $\lS H<\pi\sqrt6\,gq\MP$, which is satisfied for every $gq$ for which $\lS>0$.

The criterion~\eqref{eq:pair_criterion} can also be obtained by following the original argument of~\cite{Montero:2019ekk} in time rather than in a steady state. The pairs created at the rate $\Gamma$ separate at the speed of light, so that after a time $t$ they have built up the polarization $gq\Gamma t^2$ and the field has dropped to $E(t)=\EN-gq\Gamma t^2$. The field is removed in $t_{\rm dis}=(\EN/gq\Gamma)^{1/2}$, and the condition $t_{\rm dis}<H^{-1}$ coincides with eq.~\eqref{eq:pair_criterion} up to $\ln2$ in $\lS$. The exponential dependence of $\Gamma$ on the field does not stop the discharge: the creation rate is quenched once the field has dropped by the fraction $1/S$, but the pairs created until then are not consumed and keep carrying the current $J=2gq\Gamma t_{\rm q}$, with $t_{\rm q}=t_{\rm dis}/\sqrt S$, so that the whole field is removed in $t_f\simeq(S/4)^{1/2}\,t_{\rm dis}$. This replaces $\lS$ by $\lS-\ln(\lS/2)$, a $3\%$ change for the electron charge that we neglect. With the powers of $gq$ counted in four dimensions, i.e.\ including the modes that are not spherically symmetric on the $S^2$ (Appendix~\ref{app:discharge}), $t_{\rm dis}$ is the discharge time estimated in~\cite{Montero:2019ekk}.

\paragraph{Finite temperature.} Above $\Tc$ the exponent $S_T$ of eq.~\eqref{eq:Scwhinger-T} decreases monotonically with $T$. For $T\gg\Tc$ it reduces to the Boltzmann factor of a pair at rest (Appendix~\ref{app:instanton}),
\begin{equation}\label{eq:HighT_correct}
    S_T=\frac{2m}{T}\left(1-\frac{\Tc^2}{6T^2}+\mathcal O\!\left(\frac{\Tc^4}{T^4}\right)\right)
    =\frac{m}{T}\left(2-\frac{(gqE)^2}{12\,m^2T^2}+\ldots\right),
\end{equation}
and the process is physically the creation of a pair out of the thermal photons, which the field then separates. We keep $S_T$ as the exponent and take the prefactor from the same process at zero field, which is photon fusion rather than zero-temperature pair creation. Detailed balance with the annihilation cross section $\langle\sigma v\rangle=\pi\alpha_Q^2/m^2$ of a non-relativistic pair gives $\Gamma_T=n_+n_-\langle\sigma v\rangle=\alpha_Q^2\,mT^3e^{-S_T}/2\pi^2$ for $T\gg\Tc$, with $\alpha_Q=(gq)^2/4\pi$. This is the weak-field limit of the perturbative rate of~\cite{Torgrimsson:2019sjn} (Appendix~\ref{app:discharge}). The rate therefore scales as $(gq)^4$ rather than $(gq)^2$. With $J_{\rm p}\simeq2gq\Gamma_T/H$ the criterion~\eqref{eq:criterion} becomes $S_T<\lS^{(T)}$ and the pair-creation channel is inoperative for
\begin{equation}\label{eq:pair_thermal}
    S_T(m,T)\;\geq\;\lS^{(T)}\equiv\ln\!\left[\;\frac{gq\,\alpha_Q^2\,mT^3}{\pi^2\sqrt6\,\MP H^3}\right],
\end{equation}
which for $T\gg\Tc$ gives $m\geq\tfrac12\lS^{(T)}T$ up to small corrections. Since $\Tc=gq\EN/(2m)$, the ratio $\Tc/\Tmax=gq\EN/(2m\Tmax)$ is tiny in the present-vacuum applications of Section~\ref{sec:particlephysics} ($2\times10^{-3}$ for $gq=e$ and $m=0.2$\,eV, $\lesssim3\times10^{-26}$ for gauged $B-L$), so the high-temperature form is adequate there (during inflation, Section~\ref{sec:inflation}, it can be of order one, and only the conduction channel, which does not involve $\Tc$, is used). Because of the $(gq)^5$ in the argument of the logarithm, the resulting bound decreases quickly with the coupling: at the present vacuum energy $\lS^{(T)}/2=65$ for $gq=e$ ($m_{\min}=0.16$\,eV), and the solution of eq.~\eqref{eq:pair_thermal} drops below $m\simeq10\,T$ for $gq\lesssim10^{-10}$, where the non-relativistic form of the rate ceases to apply. Below that value one would need the relativistic thermal rate, including the medium effects of the plasma, and we draw no conclusion from the photon-fusion channel there. (The formal roots of eq.~\eqref{eq:pair_thermal} disappear at $gq\simeq5\times10^{-12}$, outside the validity of the expression.) The naive criterion ``$S_T\gtrsim1$'' would instead give $m\gtrsim T/2$, and ``$S_T\gtrsim\pi$'', the choice that reproduces eq.~\eqref{eq:FLoriginal} at $T\to\Tc$, would give $m\gtrsim\pi T/2$. Both miss the logarithm, which is the dominant effect, and both would suggest a bound that survives as $gq\to0$, which the photon-fusion channel, within its range of validity, does not support. The bound does survive to far smaller couplings, down to $gq\sim10^{-30}$, but for a different reason, to which we now turn.

\subsection{Conduction by the thermal population}\label{sec:conduction}

In an equilibrium state of temperature $T$ the charged species is present with the number density
\begin{equation}\label{eq:nth}
    n=g_s\int\!\frac{\dd^3p}{(2\pi)^3}\,\frac{1}{e^{\sqrt{p^2+m^2}/T}\pm1}
    \;\xrightarrow{\;m\gg T\;}\;g_s\left(\frac{mT}{2\pi}\right)^{3/2}e^{-m/T}
\end{equation}
for particles and, separately, for antiparticles, with the upper sign for fermions and the lower one for bosons. The two densities are equal in the absence of a chemical potential. These particles do not have to be created. The electric field of the black hole accelerates particles and antiparticles in opposite directions, and they reach the speed of light after a time $t_{\rm acc}=m/(gq\EN)$. As we verify in Appendix~\ref{app:discharge}, this is shorter than the Hubble time, and shorter than the mean free time between collisions with the bath, over the whole range of parameters considered below. At the central normalization this holds down to $(gq)_{\min}$ itself. A reduced drift velocity $\bar v<1$ enters only at the upper edge of the sensitivity band of Section~\ref{sec:TFLbound}, where the current is multiplied by $e^3$, for couplings near and below $(gq)_{\min}$, and the numerical results include it there. The resulting current density is $J_{\rm c}=2gq\,n\,\bar v$ with $\bar v\simeq1$, and the criterion~\eqref{eq:criterion} becomes
\begin{equation}\label{eq:cond_criterion}
    t_{\rm c}=\frac{\EN}{2gq\,n\,\bar v}<H^{-1}
    \quad\Longleftrightarrow\quad
    2gq\,n\,\bar v>\sqrt6\,\MP H^2\,.
\end{equation}
Eq.~\eqref{eq:cond_criterion} is a screening condition. The relative displacement of the particles and the antiparticles by a distance $\delta$ builds up the counter-field $gq\,n\,\delta$, so that the field of the black hole is canceled once $\delta$ reaches the screening length $\delta_{\rm s}=\EN/(gq\,n)$ of the throat plasma. Since the two species separate at the speed of light, $\delta=2t$, and $t_{\rm c}<H^{-1}$ states that $\delta_{\rm s}$ is shorter than the distance $2/H$ that the carriers travel in a Hubble time. The carriers are not consumed. In the relativistic, frozen-geometry limit their current remains approximately saturated through the first field zero, producing a plasma oscillation of period $4t_{\rm c}$. The current reverses when the carrier momentum changes sign. In the coupled geometry the density evolves, and the throat can collapse before the first field reversal, as in the $t_{\rm c}=H^{-1}$ example of Appendix~\ref{app:discharge}. In the Boltzmann regime eq.~\eqref{eq:cond_criterion} is a condition on the exponent $m/T$: rapid discharge is impossible if and only if
\begin{equation}\label{eq:TFL}
    \frac{m}{T}\;\geq\;\lc\,,\qquad
    \lc\equiv\ln\!\left[\frac{2g_s\,gq\,(mT/2\pi)^{3/2}\,\bar v}{\sqrt6\,\MP H^2}\right]\;.
\end{equation}
Eq.~\eqref{eq:TFL} is an implicit equation for the smallest allowed mass, $m_{\min}(gq,T)$. The current $2gq\,n\bar v$ decreases monotonically with $m$ at fixed $T$ (the exact density does, and $\bar v$ is non-increasing), so that the threshold is unique whenever the massless current exceeds $\sqrt6\MP H^2$ and the bound is empty otherwise, and because $m$ enters the logarithm only weakly the solution is insensitive to the details. For $m\lesssim T$ the density saturates at its relativistic value, $n\to\tfrac34\tfrac{\zeta(3)}{\pi^2}g_sT^3$ for fermions, and eq.~\eqref{eq:cond_criterion} shows that conduction can discharge the black hole only if
\begin{equation}\label{eq:gqmin}
    gq\;\geq\;(gq)_{\min}\equiv\frac{\sqrt6\,\MP H^2}{2\,n(0,T)}\simeq\frac{2\pi^2}{3\zeta(3)\,g_s}\,\frac{\sqrt6\,\MP H^2}{T^3}\,.
\end{equation}
For couplings below $(gq)_{\min}$ no thermal bound exists, because even an unsuppressed relativistic plasma of the charged species is too dilute to neutralize the black hole. For still smaller couplings the particles would in addition not even be accelerated to relativistic velocities within a Hubble time (Appendix~\ref{app:discharge}). For $V=V_0$, $T=\Tmax$ and $g_s=2$ one finds $(gq)_{\min}=1.5\times10^{-30}$. In general, $(gq)_{\min}=\frac{2\sqrt6\,\pi^2}{3\zeta(3)g_s}\big(\frac{\sigma}{3c}\big)^{3/4}\sqrt{H/\MP}$, i.e.\ $2.1\sqrt{H/\MP}$ for $g_*=2$ and $42\sqrt{H/\MP}$ for $g_*=106.75$ ($c=1$, $g_s=2$). These values refer to a bath held at the fixed temperature $\Tmax$, which does not count the energy of the charged species itself. Near the endpoint that species is relativistic and adds $7/2$ to $g_*$. Counting it lowers the temperature of the $c=1$ state to $1.9$\,meV, which raises $(gq)_{\min}$ to $3.2\times10^{-30}$ and lowers $m_{\min}$ at $gq=10^{-29}$ from $10$ to $7.6$\,meV. Both shifts lie inside the sensitivity band of Section~\ref{sec:TFLbound}, and for $gq\geq10^{-28}$ the bound changes by less than $3\%$.

Two remarks. First, the conduction channel dominates over the photon-fusion channel of Section~\ref{sec:pair} wherever the latter is valid ($T>\Tc$, $m\gg T$): comparing eq.~\eqref{eq:TFL} with the high-temperature form of eq.~\eqref{eq:pair_thermal}, $m/T\geq\lS^{(T)}/2$, one finds

\begin{equation}\label{eq:compare}
    \lc-\frac{\lS^{(T)}}{2}=\frac{3}{2}\ln\frac{1}{gq}+\ln\frac{m}{\sqrt{\EN}}+\ln\!\big(4\sqrt{2\pi}\big)\;>\;0
\end{equation}
for all $gq<1$ and $m>T$ at $T\simeq\Tmax$, where the second term is $\ln[(m/T)(c/2\sigma)^{1/4}]$ at $T=\Tmax$ and the last one is $2.3$ for a Dirac fermion, $g_s=2$ ($8.3$ in total for $gq=e$ and $m=0.18$\,eV, Appendix~\ref{app:discharge}). The comparison is between the two channels we calculate. Others, such as the worldline-sphaleron regime of~\cite{Gould:2018ovk}, we do not evaluate. Physically, the particles already present in the bath carry a current that is suppressed only by one power of $gq$, whereas pair creation is suppressed by $(gq)^5$, and their density is controlled by the single-particle Boltzmann factor $e^{-m/T}$ rather than by the pair factor $e^{-2m/T}$ of eq.~\eqref{eq:HighT_correct}. Second, thermal corrections to the mass, $m^2(T)=m_0^2+a\,g^2q^2T^2$ with $a=\mathcal O(1/8$--$1/4)$ for a Dirac fermion or a complex scalar in an Abelian plasma~\cite{Kraemmer:1994az,Carrington:2008dw}, are irrelevant here: the bound requires $m\geq\lc T\gg gqT$, so that $m(T)\simeq m_0$ to the accuracy of the logarithm. (If the mass is generated by a scalar condensate that evaporates at high temperature, as for the Standard Model fermions above the electroweak crossover temperature $T_{\rm EW}\simeq159$\,GeV~\cite{DOnofrio:2015gop}, then $m_0=0$ and, for $gq>(gq)_{\min}$, the bound cannot be satisfied, the thermal self-energy $\sim gqT$ being far below $\lc T$. The exception is the immediate vicinity of the endpoint, where $\lc$ itself drops to zero: there a thermal mass, which lowers the density as well, can make the criterion satisfiable in a narrow interval of couplings that we have not resolved. This is the situation discussed in Section~\ref{sec:SMEFT} and Appendix~\ref{sec:reheating}.)

\subsection{The thermal Festina Lente bound}\label{sec:TFLbound}

\begin{figure}[t]
\centering
\includegraphics[width=0.72\textwidth]{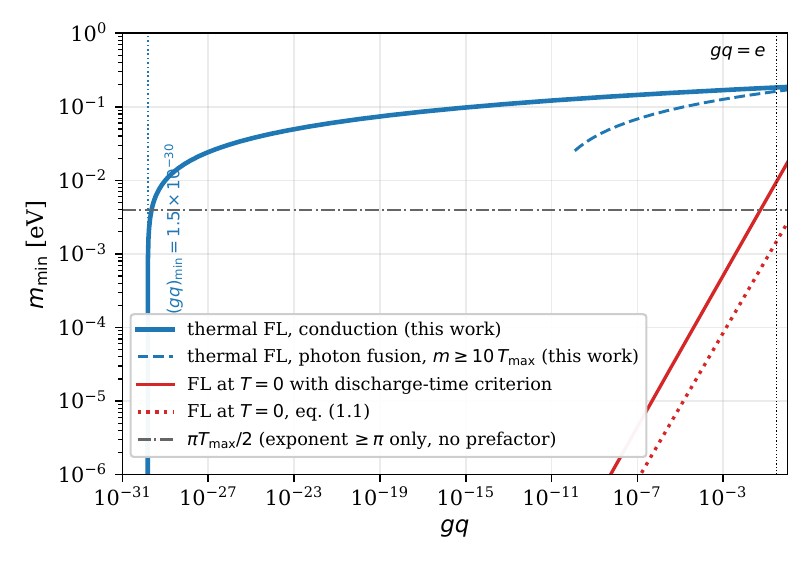}
\caption{Lower bound on the mass of a particle of charge $q$ under a $U(1)$ with coupling $g$ and gauge-boson mass below the Hubble scale, at the present vacuum energy $V_0=(2.24\,{\rm meV})^4$, as a function of $gq$. Solid blue: thermal FL bound~\eqref{eq:TFL_final} from conduction, for $\Tmax=(V_0/\sigma)^{1/4}=2.5$\,meV ($c=1$, $g_*=2$, a fixed bath and ballistic drift $\bar v$, i.e.\ the formal envelope of Section~\ref{sec:criterion}) and a Dirac fermion ($g_s=2$). It exists for $gq\geq(gq)_{\min}=1.5\times10^{-30}$ and falls to zero there. Dashed blue: thermally assisted pair creation, eq.~\eqref{eq:pair_thermal} with the photon-fusion rate, shown where its solution satisfies $m>10\,T$ ($gq\gtrsim10^{-10}$). Below this the non-relativistic rate does not apply and no curve is drawn. It is weaker than the conduction bound wherever it is valid. Red: zero-temperature FL bound with the discharge-time criterion, eq.~\eqref{eq:FL_log} (solid), and in the conventional form~\eqref{eq:FLoriginal} (dotted). The dash-dotted line is $\pi\Tmax/2$, the value obtained if the logarithms are ignored.}
\label{fig:massbound}
\end{figure}

Collecting the results, the thermal Festina Lente conjecture of Section~\ref{sec:criterion} implies, for every particle charged under a $U(1)$ whose gauge boson has a mass $m_A<H$, and with $gq\geq(gq)_{\min}$,
\begin{equation}\label{eq:TFL_final}
    m\;\geq\;\lc\!\left(gq,m,\Tmax\right)\,\Tmax\,,\qquad \Tmax=\left(\frac{cV}{\sigma}\right)^{1/4},
\end{equation}
with $\lc$ defined in eq.~\eqref{eq:TFL}. For $gq<(gq)_{\min}$ only the zero-temperature bound~\eqref{eq:FL_log} applies, and it is empty for $gq<2.3\times10^{-20}$ at the present vacuum energy (the photon-fusion channel is evaluated only for $gq\gtrsim10^{-10}$). The bound~\eqref{eq:TFL_final} has the following properties.
\begin{itemize}
\item It is of order $V^{1/4}$ up to a logarithm. The charge and the gauge coupling enter only through $\ln(gq)$: at $V=V_0$, $g_s=2$, $c=1$, $g_*=2$ (so that $\Tmax=2.5$\,meV) one has $\lc=74$ for $gq=e$ and $\lc=17$ for $gq=10^{-24}$, i.e.\ $m_{\min}=0.18$\,eV and $0.043$\,eV respectively (Table~\ref{tab:mmin}).
\item At the present vacuum energy it is stronger than the zero-temperature FL bound~\eqref{eq:FL_log} for all $gq\leq1$ (Figure~\ref{fig:massbound}): eq.~\eqref{eq:FL_log} scales as $(gq)^{1/2}V^{1/4}$ and gives $9.5$\,meV for $gq=e$, whereas eq.~\eqref{eq:TFL_final} gives $0.18$\,eV and decreases only logarithmically with $gq$.
\item The unknown order-one parameter $c$ and the content of the bath enter only through $\Tmax\propto(c/g_*)^{1/4}$ and, logarithmically, through $\lc$. The endpoint $(gq)_{\min}$ scales as $(g_*/c)^{3/4}$, eq.~\eqref{eq:gqmin}.
\item Its derivation relies on the semiclassical description of the discharge, which requires $m\gg H$ and $m\gg T$. Both are satisfied by the solution, $m_{\min}=\lc\Tmax\gg\Tmax\gg H$, whenever $\lc\gg1$, i.e.\ except in the immediate vicinity of $(gq)_{\min}$.
\end{itemize}
Figure~\ref{fig:massbound} shows $m_{\min}$ as a function of $gq$ for the present vacuum energy, together with the pair-creation bounds. Table~\ref{tab:mmin} lists representative values.

\paragraph{Fate of the carriers.} The criterion~\eqref{eq:criterion} is a statement about the charge of the black hole: it is met once the carriers have removed an order-one fraction of that charge, and with it an order-one fraction of the field energy, which, as discussed in Section~\ref{sec:criterion}, initiates the contraction of the throat. One may nevertheless ask whether the carriers, which at this point form two cold counter-streaming beams, also isotropize into the ``radiation cloud'' of~\cite{Montero:2019ekk}, and whether a criterion that required them to do so through collisions would give a different bound. The answer to the first question is that they may, through collective instabilities rather than through collisions, but that this is not established by our estimates (Appendix~\ref{app:discharge}). At the threshold density $n_{\rm th}=\sqrt6\MP H^2/(2gq)$ the plasma frequency is $\omega_p^2=2(gq)^2n_{\rm th}/m=H/t_{\rm acc}\gg H^2$, and two counter-streaming beams are unstable to the Weibel (filamentation) instability~\cite{Weibel:1959zz,Medvedev:1999tu,Bret:2010ifn}, which grows at the rate $\omega_p/\sqrt\gamma$. Integrated over the first discharge, during which $\gamma=t/t_{\rm acc}$, this gives $\int_{t_{\rm acc}}^{t_{\rm c}}\Gamma_W\dd t=2\bigl(1-\sqrt{t_{\rm acc}/t_{\rm c}}\bigr)\leq2$ $e$-folds at any density. That amplifies the seed fluctuations by less than an order of magnitude, not enough by itself to reach the nonlinear stage. What happens during the subsequent field reversals would require a growth calculation on the oscillating background, which we do not attempt. The beams are therefore susceptible to collective isotropization, but the ``radiation cloud'' of~\cite{Montero:2019ekk} is not a result of this paper, and the bound does not depend on it, since the collapse does not. Binary collisions and radiative losses, on the other hand, are negligible: the carriers reach $\gamma\gg1$ within $t_{\rm acc}\ll t_{\rm c}$, the annihilation and large-angle scattering cross sections then fall as $\alpha_Q^2/(\gamma^2m^2)$, and the optical depth of a carrier for these processes on the other carriers is
\begin{equation}\label{eq:taubin}
    \tau_{\rm bin}\simeq n_{\rm th}\,\sigma_T\,t_{\rm acc}=\frac{(gq)^2}{12\pi}\,\frac{H}{m}\,,
\end{equation}
with $\sigma_T=8\pi\alpha_Q^2/3m^2$, i.e.\ $10^{-35}$ for the electron charge and $10^{-81}$ for $g_{B-L}=10^{-24}$. Bremsstrahlung on the other carriers, whose cross section $\sim\alpha_Q^3/m^2$ does not fall with $\gamma$, removes a fraction of order $\alpha_Q^3\EN/(gq\,m^2)$ of the energy, $7\times10^{-10}$ for the electron charge up to a logarithm of order ten (Appendix~\ref{app:discharge}). The energy radiated on the photons of the bath is about $0.5\%$ of the energy gained from the field, eq.~\eqref{eq:fKN}. A criterion based on collisional thermalization would therefore be a very different criterion, and we do not adopt it. It would require a carrier density larger than $n_{\rm th}$ by the factor $1/\tau_{\rm bin}$, i.e.\ a logarithm larger by $\ln[12\pi m/(gq)^2H]\simeq80$--$210$. No equilibrium population with $T\leq\Tmax$ can provide that at the masses where the bound is evaluated, $m\gtrsim10\,T$, since there $n(m,T)\leq n(0,\Tmax)\ll n_{\rm th}/\tau_{\rm bin}$ for every $gq\leq1$. Such a criterion would remove every thermal bound of Table~\ref{tab:mmin}. Applied to Schwinger pairs in its simplest form, with one collision per Hubble time, the low-energy cross section $\sigma\sim\alpha_Q^2/m^2$ and the pair density $\Gamma/H$, i.e.\ ignoring the acceleration, it reads $\Gamma\sigma_T/H>H$ and replaces $\lS$ in eq.~\eqref{eq:FL_log} by $\lS-\ln[12\pi m^2/(gq)^3\EN]$. This changes the bound by $4\%$ for the electron charge today, but it removes the zero-temperature bound for $gq\lesssim3\,(m/\MP)^{1/3}\simeq10^{-10}$ instead of $2\times10^{-20}$. With the acceleration to $\gamma\gg1$ included, eq.~\eqref{eq:taubin}, the replacement is instead $\lS\to\lS-\ln[12\pi m/(gq)^2H]$, from $132$ to $55$ for the electron charge, and the bound is removed for $gq\lesssim5\times10^{-6}$. At finite temperature the corresponding condition, that a carrier scatter on the photons of the bath within a Hubble time, $n_\gamma\sigma_T>H$, is met only for $gq\gtrsim5\times10^{-7}$ (for $m\sim0.1$\,eV), and where it is met the energy transferred is the fraction $f_{\rm KN}\simeq0.5\%$ of eq.~\eqref{eq:fKN}: the number of scatterings is not a measure of the energy converted. The collapse of the throat does not wait for the collisions: it follows from the conversion of the field energy, and that is what fixes the bound.

\paragraph{Accuracy of the bound.} Eq.~\eqref{eq:TFL_final} is derived at the logarithmic level: the physics fixes the argument of $\lc$ up to order-one factors, each of which enters $\lc$ (and $\lS$) additively. There are four of them. The first is the fraction $f$ of the field that must be removed for the throat to collapse: $-\ln f$, i.e.\ $+0.7$ for $f=1/2$, or $+1.2$ if half of the field \emph{energy} is meant, $f=1-\sqrt{1/2}$ (a smaller fraction is removed sooner, and the bound tightens). The second is the number $\kappa$ of Hubble times allowed for the conversion: $\ln\kappa$, or $\pm1.1$ for $1/3<\kappa<3$, a range that includes the lifetime of the throat in the bath for $c\gtrsim0.02$ (Appendix~\ref{app:discharge}). The third is the prefactor of the current, that is, the velocity factor $\bar v$, the flat-space form of the density in a throat of curvature $H\ll T$, and the transverse structure of the discharge ($\pm1$). The fourth, which affects $\lS$ only, is the accumulation of the pairs discussed in Section~\ref{sec:pair} ($-\ln(\lS/2)\simeq-4$ for the electron charge). We therefore quote, as a sensitivity band, the variation of the results when the current in eq.~\eqref{eq:cond_criterion} is multiplied by $e^{\pm3}$ and the criterion is re-solved. Wherever the Boltzmann form applies this is $\delta\lc=\pm3$ and, asymptotically for $\lc\gg1$, $\delta m_{\min}\simeq\pm3\Tmax$: $\pm4\%$ for the electron charge ($\lc=74$) and $\pm19\%$ for $gq=10^{-24}$ ($35$--$51$\,meV instead of $43$\,meV). The solved intervals widen toward the endpoint ($7$--$26$\,meV instead of $17$\,meV at $gq=10^{-28}$), where $\lc$ itself is of order one and the ballistic description is at the edge of its validity. The endpoint $(gq)_{\min}$ is not protected by a logarithm at all: it moves between $3\times10^{-31}$ and $3\times10^{-29}$, so that at $gq=10^{-29}$ the bound ranges from zero to $20$\,meV. This band is a specified variation of the normalization of the current, not a total error budget: it does not cover the assumptions of Section~\ref{sec:criterion} about the geometry, nor the choice of the bath, which enters through $\Tmax\propto(c/g_*)^{1/4}$ and changes the bound itself (Section~\ref{sec:neutrinos}). It is quoted for the benchmarks of Section~\ref{sec:particlephysics}, with $\lc\gg1$. Where the logarithm is of order a few, as for the zero-temperature bound near the upper end of the inflationary range of Section~\ref{sec:slowroll} ($\lS\simeq4$), the same variation multiplies the threshold mass by $0.51$--$1.32$ instead of shifting it by a few percent. The thermal bound is more sensitive still there, since $\lc$ itself is only of order one ($1.8$ at $H_I=4.7\times10^{13}$\,GeV for $c=1$, Appendix~\ref{sec:reheating}). In Section~\ref{sec:particlephysics} we quote these sensitivities wherever a comparison with data depends on them. The central values are for $f=1$, $\kappa=1$, $\bar v$ as in Appendix~\ref{app:discharge}, $c=1$ and $g_*=2$.

\begin{table}[t]
\centering
\begin{adjustbox}{max width=\linewidth}\begin{tabular}{lcccccccc}
\toprule
$gq$ & $e$ & $10^{-3}e$ & $10^{-10}e$ & $10^{-15}e$ & $10^{-21}e$ & $10^{-24}$ & $10^{-26}$ & $10^{-28}$\\
\midrule
$\lc$ & 74 & 67& 50 & 38 & 24 & 17 & 12 & 7.0\\
$m_{\min}$ [eV] & 0.18 & 0.17 & 0.12 & 0.095 & 0.059 & 0.043 & 0.031 & 0.017\\
\bottomrule
\end{tabular}\end{adjustbox}
\caption{The thermal FL bound~\eqref{eq:TFL_final} at the present vacuum energy for representative values of $gq$ ($c=1$, $g_*=2$, $g_s=2$). The exact thermal integral~\eqref{eq:nth} and the velocity factor $\bar v$ of Appendix~\ref{app:discharge} are used. The bound scales as $(c/g_*)^{1/4}$ up to the logarithm. For a complex scalar ($g_s=1$) $\lc$ decreases by $\ln2$ in the Boltzmann regime (near the endpoint the Bose density must be used). The sensitivity to the normalization of the current (Section~\ref{sec:TFLbound}) is $\delta\lc=\pm3$, i.e.\ about $\pm7.5$\,meV in $m_{\min}$ for the first entries and wider intervals near the endpoint ($7$--$26$\,meV at $10^{-28}$).}
\label{tab:mmin}
\end{table}

\section{Particle physics constraints}
\label{sec:particlephysics}
The thermal FL bound~\eqref{eq:TFL_final} applies to any gauged $U(1)$ interaction in a dS space whose gauge boson is lighter than the Hubble scale, and consequently could be used to set constraints on scenarios where such interactions arise. In this section, we discuss several phenomenological implications. Throughout, ``today'' refers to $V=V_0=(2.24\,{\rm meV})^4$ and $H_\Lambda=1.2\times10^{-33}$\,eV. In Sections~\ref{sec:neutrinos} and~\ref{sec:mCP} the numbers are quoted for $c=1$, $g_*=2$ ($\Tmax=2.5$\,meV) and Dirac fermions ($g_s=2$). For other choices the bounds scale as $(c/g_*)^{1/4}$, up to the logarithm.

\subsection{Neutrinos}\label{sec:neutrinos}

It is a remarkable coincidence that the FL bound is set by the vacuum energy on the one hand, while on the other hand for the vacuum energy of today, the mass scale of neutrinos is roughly given by $m^{4}_{\nu}\sim V_{0}$. This seems to suggest a possible link between the FL bound and neutrino masses, which was explored in~\cite{Montero:2021otb}. However, such a link runs into trouble as neutrinos are known to be neutral. Nonetheless, neutrinos are charged under the global $B-L$ symmetry\footnote{Here we assume that neutrinos are Dirac fermions. Neutrinos are also charged under $L_{\tau}-L_{\mu}$, $L_{\tau}-L_{e}$, and $L_{\mu}-L_{e}$. In the SM extended by three right-handed neutrinos, any linear combination of $B-L$, $L_i-L_j$ and hypercharge can be gauged without anomalies, but an unbroken flavor-dependent combination with charge matrix $Q_L$ would have to satisfy $[Q_L,M_\nu M_\nu^\dagger]=0$ (equivalently $Q_LM_\nu=M_\nu Q_R$), which the observed mixing forbids. The case considered here is $B-L$, under which all three mass eigenstates carry unit charge.}. Global symmetries are generally believed to be forbidden in a consistent theory of quantum gravity~\cite{Kallosh:1995hi, Banks:2010zn}, which implies that $B-L$ cannot be a true global symmetry. This leaves three possibilities. Either $B-L$ is an accidental symmetry broken explicitly at some high scale, or it is gauged and spontaneously broken, with a gauge boson much heavier than $H_\Lambda$ (this is what happens in the seesaw mechanism~\cite{Minkowski:1977sc,GellMann:1980vs,Yanagida:1979as,Mohapatra:1979ia}, where the Majorana mass of the right-handed neutrinos breaks $B-L$), or it is gauged and unbroken, with Dirac neutrinos and a massless (or ultralight) gauge boson. Only the last possibility is constrained by the FL argument, and we assume it in this subsection. In light of the FL bound, and under the assumption that the $B-L$ force carrier satisfies $m_{B-L} < H_\Lambda$, this last possibility sets the following bound on the masses of neutrinos:
\begin{equation}
    \label{eq:FLC_neutrono_mass}
    m_{\nu}^{4} \gtrsim 2 g_{B-L}^{2}V\,.
\end{equation}
However, given that fifth-force searches and equivalence-principle tests limit $g_{B-L}\lesssim10^{-24}$ in the mass range $m_{B-L} \lesssim H_\Lambda$~\cite{Heeck:2014zfa}, the FL bound does not yield a meaningful bound ($m_\nu\gtrsim3\times10^{-15}$\,eV). With the discharge-time criterion of Section~\ref{sec:pair} it is in fact empty, since $\lS<0$ for $g_{B-L}<2\times10^{-20}$. Things are different when the thermal bound is applied.

\begin{figure}[t]
\centering
\includegraphics[width=0.7\textwidth]{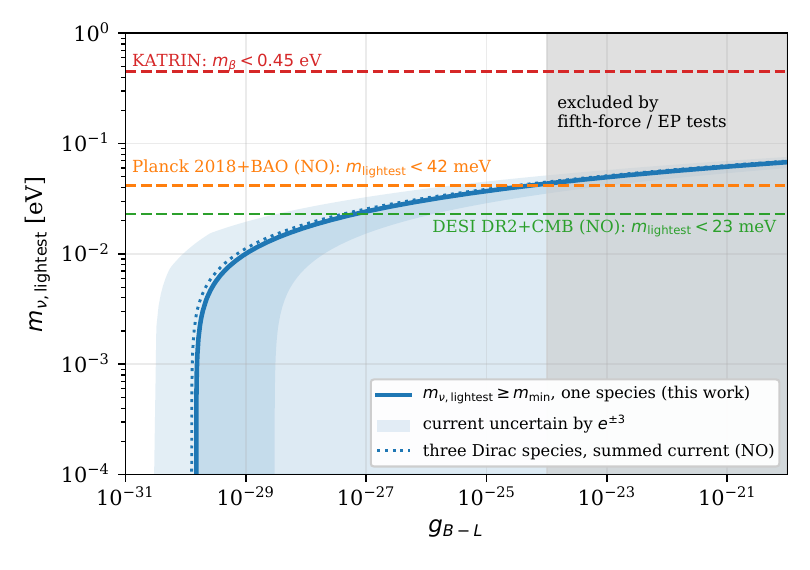}
\caption{Lower bound on the mass of the lightest neutrino, eq.~\eqref{eq:nu_bound}, for an unbroken gauged $B-L$ with $m_{B-L}<H_\Lambda$ and Dirac neutrinos, as a function of $g_{B-L}$ (blue). The shaded region below the curve is excluded by the thermal FL bound. The horizontal lines show the upper limits on $m_{\nu,{\rm lightest}}$ from KATRIN~\cite{KATRIN:2024cdt}, from Planck\,2018+BAO~\cite{RoyChoudhury:2019hls} and from DESI\,DR2\,BAO+CMB with the oscillation constraints imposed~\cite{Elbers:2025vlz} (normal ordering). The KATRIN limit is on the effective mass $m_\beta$. The gray band is excluded by fifth-force and equivalence-principle tests~\cite{Heeck:2014zfa}. The solid curve uses the lightest state alone in a bath of fixed temperature $\Tmax$ ($c=1$, the formal envelope of Section~\ref{sec:criterion}, $g_*=2$, ballistic drift). The dotted curve sums the currents of the three Dirac mass eigenstates for the normal ordering at the same fixed temperature (not a common energy-budget solution, Section~\ref{sec:neutrinos}). The light-blue band is the sensitivity of the bound to a multiplicative factor $e^{\pm3}$ in the current (Section~\ref{sec:TFLbound}). Where it reaches the bottom of the plot the bound may be absent.}
\label{fig:neutrino_gBL}
\end{figure}

All three neutrino mass eigenstates carry unit $B-L$ charge and, in the maximal bath, are part of the equilibrium plasma, so eq.~\eqref{eq:TFL_final} must hold for each of them. The lightest one is the weakest link:
\begin{equation}\label{eq:nu_bound}
    m_{\nu,{\rm lightest}}\;\geq\;m_{\min}(g_{B-L})\simeq\lc\,\Tmax\,,\qquad \lc=\ln\!\left[\frac{4g_{B-L}\,(m_{\nu}\Tmax/2\pi)^{3/2}}{\sqrt6\,\MP H_\Lambda^2}\right],
\end{equation}
which is shown in Figure~\ref{fig:neutrino_gBL}. The closed form on the right is the non-relativistic, saturated-drift approximation, which is valid for $m\gg T$. All the curves and numbers quoted here and in the figures come instead from the exact Fermi--Dirac current, which departs from that form near the endpoint, where the bound falls below the temperature. Because the coupling enters only logarithmically, the bound is only weakly charge dependent away from that endpoint: $m_{\nu,{\rm lightest}}\geq43$\,meV for $g_{B-L}=10^{-24}$, $31$\,meV for $10^{-26}$, $17$\,meV for $10^{-28}$ and $10$\,meV for $10^{-29}$, with the sensitivity band of Section~\ref{sec:TFLbound} giving $35$--$51$, $22$--$39$, $7$--$26$ and $0$--$20$\,meV in turn. The bound disappears only below $(gq)_{\min}=1.5\times10^{-30}$ (between $3\times10^{-31}$ and $3\times10^{-29}$ within the same band), where a $B-L$ plasma can no longer discharge the black hole. Had we ignored the logarithm, i.e.\ required the pair-creation exponent to satisfy $2m/T\gtrsim\pi$, the criterion that reproduces eq.~\eqref{eq:FLoriginal} at zero temperature, we would have obtained $m_{\nu,{\rm lightest}}\gtrsim\pi\Tmax/2=3.9$\,meV instead, independently of $g_{B-L}$. The difference between the two shows that the value of an FL-type bound is set by the discharge criterion, and not by the exponent alone.

Experimental searches set an upper bound on neutrino masses. A conservative model-independent upper bound on the masses of neutrinos of $m_{\nu} < 0.45$\,eV can be obtained from the KATRIN experiment~\cite{KATRIN:2024cdt} regardless of the neutrino mass ordering\footnote{Strictly speaking, KATRIN limits $m_{\beta} < 0.45$\,eV at $90\%$ CL, where $m_{\beta}^{2} = \sum_{i}|U_{ei}|^{2}m_{i}^2\geq m_{\nu,{\rm lightest}}^2$.}. On the other hand, under minimal $\Lambda$CDM, DESI DR2 and CMB measurements constrain $\sum m_{\nu_i} < 0.064$\,eV at $95\%$ CL~\cite{Elbers:2025vlz,DESI:2025zgx}, which seems to disfavor the inverted mass ordering. Under the normal mass ordering the masses are
\begin{equation}
    \label{eq:normal_ordering}
    m_{2} = \sqrt{m_{1}^{2}+\Delta m_{21}^{2}}\,, \hspace{5mm} m_{3} = \sqrt{m_{1}^{2}+\Delta m_{31}^{2}}\,,
\end{equation}
and given that $\Delta m_{21}^{2}\simeq7.49\times10^{-5}$\,eV$^2$ and $\Delta m_{31}^{2}\simeq2.5\times10^{-3}$\,eV$^2$~\cite{Esteban:2024eli}, the sum limit would correspond to $m_{\nu,{\rm lightest}}<4.2$\,meV if it were a hard cut. The limit on the lightest mass obtained by the DESI collaboration from the same data, with the oscillation constraints imposed, is however considerably weaker, $m_{\nu,{\rm lightest}}<23$\,meV at $95\%$ CL~\cite{Elbers:2025vlz}: the posterior of $\sum m_\nu$ peaks below the oscillation floor of $0.059$\,eV, and the two limits are not related by a change of variables. We use the direct limit $m_{\nu,{\rm lightest}}<23$\,meV.

If we use instead the more conservative limits suggested by the Planck 2018 data combined with BAO~\cite{Planck:2018vyg, RoyChoudhury:2019hls}, $\sum m_\nu<0.15$\,eV for the normal and $0.17$\,eV for the inverted ordering, the upper bound on the lightest mass relaxes to $42$\,meV and $41$\,meV, respectively.

Combining eq.~\eqref{eq:nu_bound} with these limits (Figure~\ref{fig:neutrino_gBL}) leads to the following conclusions, all conditional on the thermal FL conjecture:
\begin{itemize}
\item If the DESI\,DR2 limit $m_{\nu,{\rm lightest}}<23$\,meV holds, an unbroken gauged $B-L$ with $m_{B-L}<H_\Lambda$ and Dirac neutrinos is excluded for all $g_{B-L}>7\times10^{-28}$ ($3\times10^{-29}$--$1.3\times10^{-26}$ for a current uncertain by $e^{\pm3}$), i.e.\ over most of the range $g_{B-L}>(gq)_{\min}=1.5\times10^{-30}$ in which the thermal bound exists. The crossing refers to the benchmark state of this paper: the lightest state alone, in a bath consisting of the two polarizations of the $B-L$ gauge boson at the fixed temperature $\Tmax$ ($c=1$, $g_*=2$). Summing the currents of the three mass eigenstates, whose masses follow from $m_{\nu,{\rm lightest}}$ and the splittings, moves the crossing to $4\times10^{-28}$. A bath that also contains the photon and the three neutrinos in equilibrium, sharing the energy budget $V_0$, has a lower temperature ($2.1$\,meV) and moves it to $3\times10^{-27}$. These numbers define the model dependence of the exclusion within the $c=1$ envelope. Let us stress again that $c=1$ is the largest value for which our bound could plausibly hold, leading to the strongest possible constraint. However, there has been no top-down derivation of which value of $c$ is correct. The bath itself matters more, because $\Tmax\propto c^{1/4}$ enters the bound and not merely its normalization. Repeating the calculation with the same single-species current at $c=0.1$ gives $\Tmax=1.40$\,meV and $m_{\min}(10^{-24})=21.8$\,meV, which is below the DESI limit. The exclusion then starts only at $g_{B-L}=2.2\times10^{-24}$, above the laboratory limit, and at $c=10^{-2}$ it starts at $2\times10^{-18}$. The exclusion quoted here is therefore a statement about the formal $c=1$ benchmark. At the upper edge of the normalization band, $e^3J$, the $c=0.1$ crossing moves to $1.1\times10^{-25}$, below the laboratory limit, while for $c=10^{-2}$ nothing laboratory-allowed is excluded anywhere in that band. Since $g_{B-L}\geq\sqrt{3/2}\,H_\Lambda/\MP\simeq6\times10^{-61}$ is required by the magnetic FL bound~\cite{Montero:2021otb}, for the two-polarization fixed-temperature benchmark only couplings in the range $6\times10^{-61}\lesssim g_{B-L}\lesssim7\times10^{-28}$ would survive, and up to $1.3\times10^{-26}$ within the sensitivity band. Below $1.5\times10^{-30}$ they survive because a $B-L$ plasma is too dilute to affect a Nariai black hole, and above that value because the bound stays below $23$\,meV.
\item With the more conservative limit $\sum m_\nu<0.15$\,eV (normal ordering), the bound $m_{\nu,{\rm lightest}}<42$\,meV is violated for $g_{B-L}>6\times10^{-25}$ for the central value of $\lc$ and $c=1$ ($7\times10^{-19}$ for $c=0.1$, far above the laboratory limit). With the current uncertain by $e^{\pm3}$ the crossing moves between $3\times10^{-26}$ and $10^{-23}$, i.e.\ it straddles the laboratory limit $g_{B-L}\lesssim10^{-24}$~\cite{Heeck:2014zfa}. This conclusion is therefore marginal: the Planck+BAO limit constrains unbroken gauged $B-L$ to couplings comparable with, or somewhat below, the laboratory limit, and a sharper statement would require the order-one factors of Section~\ref{sec:TFLbound} to be determined.
\item Conversely, if $B-L$ is an unbroken gauge symmetry with $10^{-29}\lesssim g_{B-L}\lesssim10^{-24}$, and for the same $c=1$, $g_*=2$ benchmark, the lightest neutrino must be heavier than $10$--$43$\,meV ($0$--$51$\,meV once the uncertainty of the current is included, the lower value reflecting that at $g_{B-L}=10^{-29}$ the bound may be absent altogether), i.e.\ $\sum m_\nu\gtrsim0.07$--$0.15$\,eV ($0.06$--$0.18$\,eV within the band). This prediction is within reach of the next generation of cosmological surveys.
\end{itemize}
The KATRIN limit alone does not constrain the scenario, since $m_{\min}(g_{B-L})\leq0.05$\,eV over the whole allowed range $g_{B-L}\lesssim10^{-24}$. We also note that the cosmological limits are model dependent (they assume $\Lambda$CDM and a positive-mass prior), which is why we quote the consequences for both the DESI and the Planck+BAO values.

\subsection{Millicharged particles (mCPs)}\label{sec:mCP}

Perhaps the most obvious application of the thermal bound is to set bounds on mCPs, i.e.\ particles of charge $q=\epsilon e$ under electromagnetism (we use $gq\to\epsilon e$ throughout). A simple application of the FL bound in eq.~\eqref{eq:FLoriginal} can be used to set strong bounds on the $m$--$\epsilon$ parameter space~\cite{Ban:2022jgm,Montero:2022jrc}, $m\geq(2\epsilon^2e^2V_0)^{1/4}=1.5\,{\rm meV}\times\sqrt\epsilon$. Ref.~\cite{Montero:2022jrc} also derives the corresponding constraints on dark-sector particles charged under a dark $U(1)$ and on kinetically mixed dark photons.

\begin{figure}[t]
\centering
\includegraphics[width=0.82\textwidth]{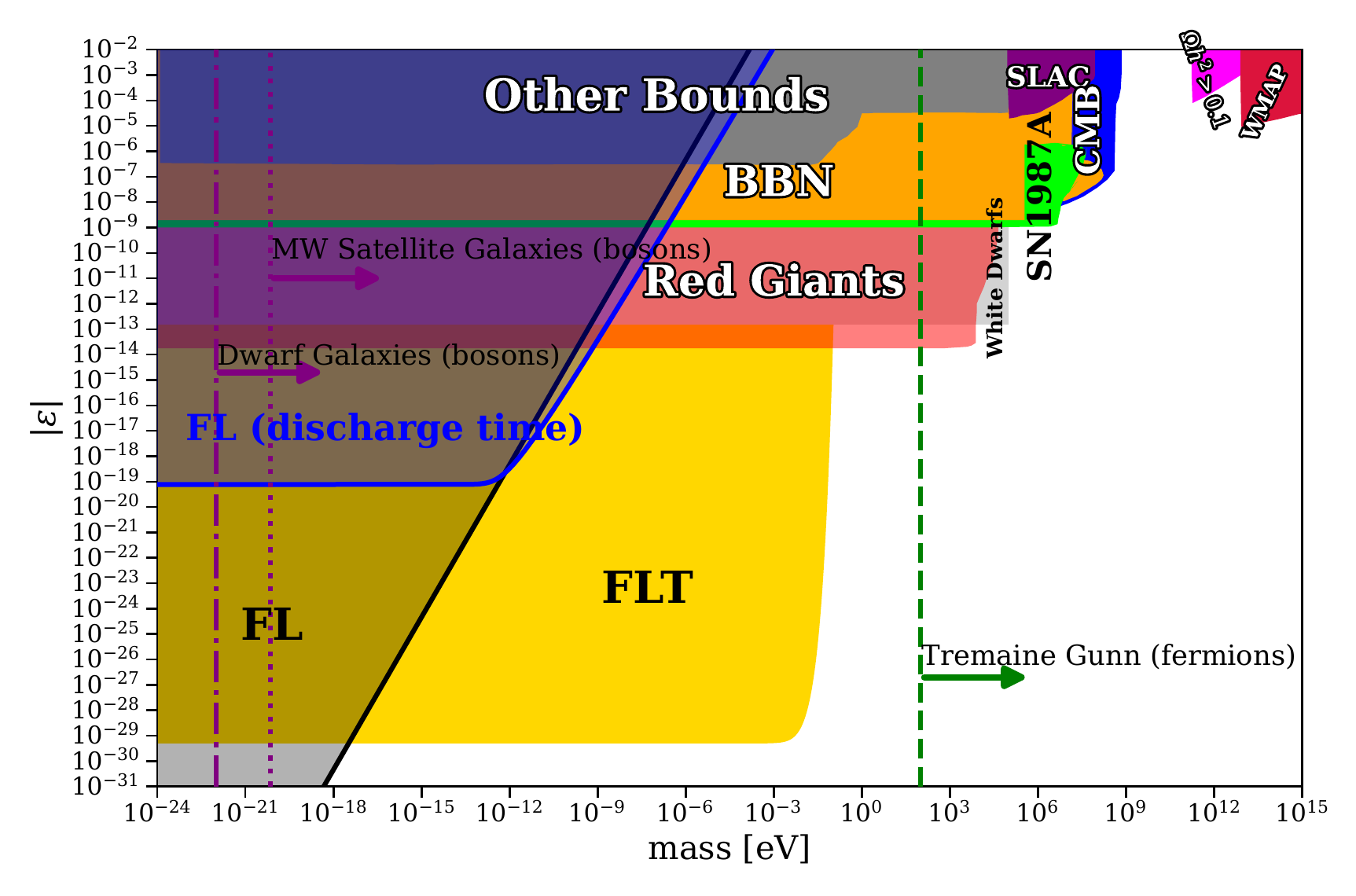}
\caption{Bounds on millicharged particles of charge $\epsilon e$ at the present vacuum energy. The region labeled FLT (gold) is excluded by the thermal FL bound~\eqref{eq:TFL_final} ($c=1$, $g_*=2$, Dirac fermion) and ends at $\epsilon=(gq)_{\min}/e=5\times10^{-30}$. The region labeled FL (olive) is excluded by the zero-temperature FL bound in the conventional form~\eqref{eq:FLoriginal} used in~\cite{Ban:2022jgm}, and the blue curve is the zero-temperature bound with the discharge-time criterion, eq.~\eqref{eq:FL_log}, which exists only for $\epsilon>8\times10^{-20}$. The other shaded regions are the accelerator (SLAC), cosmological (BBN, CMB, WMAP, $\Omega h^2>0.1$), stellar (red giants, white dwarfs, SN\,1987A) and other limits compiled in~\cite{Abu-Ajamieh:2025sul}. The vertical lines with arrows are lower limits on the dark-matter mass, from dwarf and Milky-Way satellite galaxies for bosonic dark matter (purple) and from the Tremaine--Gunn bound~\cite{Tremaine:1979we} for fermionic dark matter (green, dashed). The three theory curves are computed from eqs.~\eqref{eq:FLoriginal}, \eqref{eq:FL_log} and~\eqref{eq:TFL_final} ($c=1$, fixed bath, one Dirac species). The logarithmic curve ends at $\epsilon=7.6\times10^{-20}$, where $\lS=0$, and equals the conventional one at $\epsilon=2.2\times10^{-19}$. The experimental regions are graphical reproductions of the compilation of~\cite{Abu-Ajamieh:2025sul}.}
\label{fig:mcp_plane}
\end{figure}

With the discharge-time criterion the zero-temperature bound changes in a charge-dependent way: eq.~\eqref{eq:FL_log} exceeds eq.~\eqref{eq:FLoriginal} by the factor $(\lS/\pi)^{1/2}$ in the mass, which is larger than one for $\epsilon>2\times10^{-19}$ ($\lS>\pi$), smaller than one for $8\times10^{-20}<\epsilon<2\times10^{-19}$, and the bound is empty below $\epsilon e=2.3\times10^{-20}$, i.e.\ $\epsilon<8\times10^{-20}$. The thermal bound~\eqref{eq:TFL_final} is qualitatively different (Figure~\ref{fig:mcp_plane}): it depends on $\epsilon$ only logarithmically, $m\geq0.18$\,eV for $\epsilon=1$, $0.12$\,eV for $\epsilon=10^{-10}$, $0.095$\,eV for $\epsilon=10^{-15}$ and $0.059$\,eV for $\epsilon=10^{-21}$, and it persists down to $\epsilon=(gq)_{\min}/e=5\times10^{-30}$. For the strongest constraint at the $c=1$ benchmark, then, in a de Sitter universe with the present vacuum energy, any particle with an electric millicharge $\epsilon\gtrsim10^{-28}$ must be heavier than $10^{-2}$--$0.2$\,eV (Table~\ref{tab:mmin}). This excludes the entire ultralight region of the millicharged-particle parameter space, extending the zero-temperature FL exclusion by many orders of magnitude. This region is otherwise only accessible to astrophysical and cosmological probes~\cite{Davidson:2000hf,Vogel:2013raa}. For fermionic millicharged \emph{dark matter} the Tremaine--Gunn bound~\cite{Tremaine:1979we}, $m\gtrsim100$\,eV (the precise value depends on the halo considered), remains stronger. The thermal bound, however, applies to any charged state, bosonic or fermionic, stable or not, and irrespective of whether it constitutes the dark matter. The thermal bound assumes the mCP to be part of the equilibrium bath. If the millicharge arises from kinetic mixing with a massless dark photon, the particle couples to two massless gauge fields at once, with the charge vector $(\epsilon e,g_D)$ in the canonical basis. A Nariai black hole charged along that vector probes the full charge $\sqrt{\epsilon^2e^2+g_D^2}\geq\epsilon e$, so using $gq=\epsilon e$ gives a conservative necessary condition. The magnetic FL bound of~\cite{Montero:2021otb} constrains the fundamental gauge couplings rather than $\epsilon$, and we do not use it here.

It was shown in~\cite{Abu-Ajamieh:2025sul} that stronger bounds on mCPs can be obtained from the FL bound by assuming inflation. More specifically, assuming instantaneous reheating, the energy density at the end of inflation is converted promptly to radiation
\begin{equation}\label{eq:Inst_reh}
    3M_{P}^2H_{I}^{2} = \frac{\pi^{2}}{30}g_{\text{reh}}T^{4}_{\text{reh}}\,,
\end{equation}
where $g_{\text{reh}} = g_{*}$ is the number of relativistic degrees of freedom at reheating and $H_{I}$ is the Hubble scale during inflation. In order for reheating not to impact Big Bang Nucleosynthesis (BBN), we need to impose $T_{\text{reh}} > T_{\text{BBN}}$, with $T_{\rm BBN}\simeq2$--$6$\,MeV at 95\% CL, depending on the decay channel of the reheating particle and on the data combination~\cite{Hasegawa:2019jsa,Barbieri:2025moq}, which sets a lower bound on $H_{I}$,
\begin{equation}\label{eq:HI_lowerBound}
    H_{I} > \frac{\pi\sqrt{g_{\text{reh}}}T^{2}_{\text{BBN}}}{3\sqrt{10}M_{P}}\,.
\end{equation}
Inserted into eq.~\eqref{eq:FLoriginal}, this gives $m\gtrsim(2\epsilon^2e^2\sigma_{\rm reh})^{1/4}T_{\rm BBN}$~\cite{Abu-Ajamieh:2025sul}. Applying the thermal bound in the same way would be even more dramatic: at the end of inflation the bath \emph{is} the reheating plasma, $\Tmax\to T_{\rm reh}$, and eq.~\eqref{eq:TFL_final} would require every charged particle, for any $\epsilon$ above $(gq)_{\min}(T_{\rm reh})/e\simeq4\times10^{-20}$ (at $4$\,MeV), to be heavier than $\lc\,T_{\rm reh}$ with $\lc\simeq50$ for $\epsilon=1$ and a charge-dependent logarithm otherwise (Appendix~\ref{sec:reheating}). We do not use these reheating-era bounds here, because the same requirement applies to the electron, which violates it throughout the MeV--GeV range relevant to nucleosynthesis. The crossing below which the electron would satisfy it lies at $T_{\rm reh}\simeq9$\,keV, and above $T_{\rm reh}\simeq9\times10^{15}$\,GeV the criterion becomes empty rather than violated (Appendix~\ref{sec:reheating}). As discussed in Appendix~\ref{sec:validity}, this shows that the FL-type reasoning cannot be applied to the reheating bath, whose vacuum energy is not a cosmological constant. The zero-temperature bound of~\cite{Abu-Ajamieh:2025sul}, on the other hand, is a statement about the masses and charges during inflation at the lowest inflationary scale allowed by eq.~\eqref{eq:HI_lowerBound}, and this caveat does not affect it. With the discharge-time criterion of Section~\ref{sec:pair} it is empty below $\epsilon e=(2\pi^3H_I/\sqrt6\MP)^{1/3}$, i.e.\ for $\epsilon\lesssim1.4\times10^{-13}$ at the lowest scale ($T_{\rm BBN}\simeq4$\,MeV). It applies to the electron as well, which violates it in the electroweak vacuum for every $H_I$ between $H(T_{\rm BBN})$ and the upper edge $2.7\times10^{15}$\,GeV at which the criterion becomes empty (Section~\ref{sec:slowroll}). These inflationary bounds on millicharged particles therefore presuppose that this conflict is resolved while the photon stays massless and its coupling unsuppressed during inflation, which within the Standard Model requires the flattened Higgs potential of Section~\ref{sec:slowroll}. If the photon is instead Higgsed, or its coupling suppressed, during inflation, they do not apply.

\begin{figure}[t]
\centering

\includegraphics[width=0.7\textwidth]{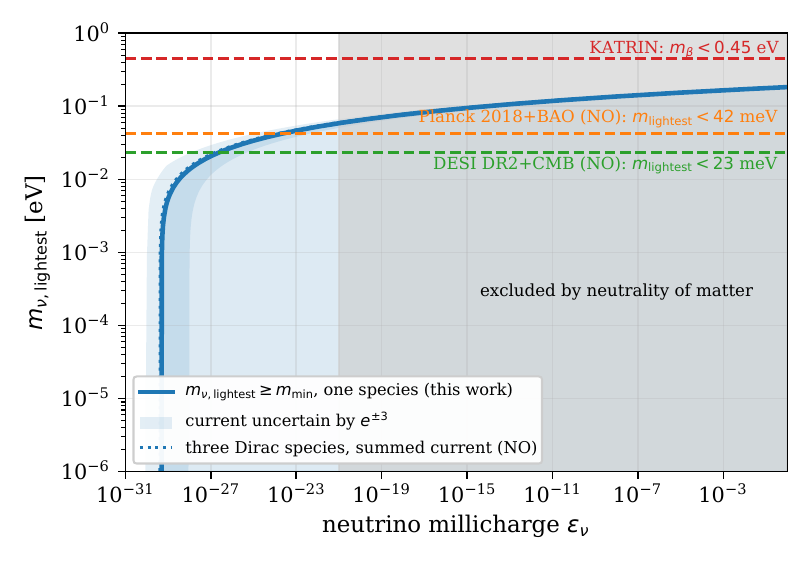}
\caption{
Thermal FL bound on the mass of the lightest neutrino as a function of a putative neutrino millicharge $\epsilon_\nu$ (blue, with the excluded region shaded), together with the upper limits on $m_{\nu,{\rm lightest}}$ from KATRIN (on $m_\beta$), Planck+BAO and DESI\,DR2\,BAO+CMB (with the oscillation constraints imposed, normal ordering), and the laboratory limit $\epsilon_\nu<10^{-21}$ from the neutrality of matter (gray), which applies to the common-charge assignment $q_n=-q_\nu$ of Section~\ref{sec:mCP}. Solid curve, dotted curve and band as in Figure~\ref{fig:neutrino_gBL}.}
\label{fig:neutrino_charge}
\end{figure}

Although in the SM neutrinos are neutral, they nonetheless could acquire an electric millicharge in several ways, including shifting the hypercharge assignment of the Standard Model by a multiple of $B-L$, $Y\to Y+2\epsilon(B-L)$ (in the convention $Q=T_3+Y/2$), introducing an extra $U(1)$ gauge group~\cite{Holdom:1985ag}, or introducing non-locality~\cite{Capolupo:2022awe, Abu-Ajamieh:2023txh, Abu-Ajamieh:2023roj, Abu-Ajamieh:2023syy, Abu-Ajamieh:2024woy,Abu-Ajamieh:2024egb}. The hypercharge shift is anomaly free once right-handed neutrinos are added, and gives Dirac neutrinos a common charge $q_\nu=-\epsilon e$ together with correlated shifts of the other charges: the neutron acquires $q_n=-q_\nu$, so that a nominally neutral atom with $N$ neutrons carries the charge $-Nq_\nu$. The neutrality-of-matter limits quoted below therefore constrain precisely this assignment~\cite{Foot:1990uf,Babu:1989ex,Okun:1983vw}. As pointed out earlier, the most conservative direct measurement bound on neutrino masses comes from the KATRIN experiment, which sets an upper bound of $m_{\nu} < 0.45$\,eV~\cite{KATRIN:2024cdt}. On the other hand, matter neutrality experiments set an upper bound of $\epsilon_{\nu} < 1 \times 10^{-21}$~\cite{Marinelli:1983nd, Raffelt:1999gv, Giunti:2014ixa, Bressi:2011yfa,Das:2020egb}.
Figure~\ref{fig:neutrino_charge} shows the thermal bound $m_{\nu,{\rm lightest}}\geq m_{\min}(\epsilon_\nu e)$ together with these limits. A neutrino millicharge is constrained far beyond the laboratory bound once the cosmological mass limits are used: with $\sum m_\nu<0.15$\,eV (Planck+BAO) one needs $\epsilon_\nu<2\times10^{-24}$ ($10^{-25}$--$4\times10^{-23}$ for a current uncertain by $e^{\pm3}$), and with the DESI\,DR2 limit $m_{\nu,{\rm lightest}}<23$\,meV one needs $\epsilon_\nu<2\times10^{-27}$ ($10^{-28}$--$4\times10^{-26}$), i.e.\ the neutrino must be neutral to better than one part in $10^{26}$ (the bound switches off only below $(gq)_{\min}/e=5\times10^{-30}$, itself uncertain by the factors of Section~\ref{sec:TFLbound}, $(1$--$100)\times10^{-30}$). These numbers belong to the $c=1$ benchmark. For $c=0.1$ the endpoint moves to $\epsilon=2.8\times10^{-29}$ and the DESI limit gives $\epsilon_\nu<7\times10^{-24}$, still below the laboratory limit $\epsilon_\nu<10^{-21}$, while for $c=0.01$ the improvement over the laboratory limit is lost. The KATRIN limit alone does not constrain $\epsilon_\nu$, since $m_{\min}\leq0.18$\,eV for all $\epsilon_\nu\leq1$.
In fact, it was shown in~\cite{Abu-Ajamieh:2024gaw} that the weak gravity conjecture~\cite{Arkani-Hamed:2006emk}, together with the experimental bounds from matter neutrality, implies that neutrinos are strictly neutral. That argument uses the magnetic form of the conjecture, $\Lambda\lesssim gq\MP$, with the cutoff $\Lambda$ inferred from collider data: it gives the \emph{lower} limit $\epsilon\gtrsim\Lambda_{\rm LEP}/(e\MP)\simeq1.4\times10^{-14}$ for $\Lambda_{\rm LEP}\simeq10$\,TeV, which is incompatible with the matter-neutrality limit quoted above, so that the millicharge has to vanish. These are additional assumptions, and we do not make them here. The existence of a superextremal state on its own, which the electron already provides for electromagnetism, does not constrain the neutrino charge. The bounds of the present section follow from the thermal FL conjecture alone, and they quantify how small a nonzero millicharge it allows. A generalized scalar version of the conjecture, applied to the Yukawa sector, is discussed in~\cite{Abu-Ajamieh:2024xic}.

\subsection{Constraining the SMEFT operators}\label{sec:SMEFT}

As pointed out in~\cite{Venken:2023hfa}, the condition in eq.~\eqref{eq:Tbound} can be used to constrain the electroweak phase transition (EWPT). See also~\cite{Lee:2021cor}. The logic is the following. In the symmetric phase the charged particles of the Standard Model are massless, up to thermal masses of order $gT$, so that neither the zero-temperature FL bound nor the thermal bound~\eqref{eq:TFL_final} can be satisfied in a vacuum-dominated symmetric phase. The universe must therefore be radiation dominated when the symmetry is restored. A sufficient diagnostic is to demand this already at the spinodal temperature $T_0$, at which the origin becomes a minimum of the thermal potential: the vacuum energy released in the transition, $V(0)$, must not exceed the radiation energy there, $\sigma T_0^4\gtrsim c\,V(0)$. The transition itself takes place at some temperature $T_n$ between the spinodal temperature $T_0$, below which the origin is no longer a minimum, and the critical temperature $T_{\rm crit}\geq T_0$, at which the two phases are degenerate. Since $\rho_R(T_n)\geq\rho_R(T_0)$, imposing the condition at $T_0$ is enough to guarantee radiation domination at the transition, and that is what we do. This is the opposite of eq.~\eqref{eq:Tbound}, which is the condition under which the FL bound \emph{would} apply, and it is a constraint on the shape of the Higgs potential. For example, consider the SM Higgs potential
\begin{equation}
    \label{eq:SM_Higgs_V}
    V(\phi) = -\mu^{2}|\phi|^{2} + \lambda|\phi|^{4} + \frac{\mu^{4}}{4\lambda}.
\end{equation}
It can be observed that the vacuum energy is given by $V(0) = \frac{\mu^{4}}{4\lambda}$. On the other hand, at the symmetry-restoration temperature, $\mu^{2} =\alpha T_{0}^{2}$, where $\alpha$ is a function of the couplings. Thus the requirement $\sigma T_0^4\geq cV(0)$ sets a lower bound on the Higgs quartic coupling
\begin{equation}
    \label{eq:lambda_min}
    \lambda\geq \frac{c\,\alpha^{2}}{4\sigma}\,,
\end{equation}
which the Standard Model satisfies comfortably: with $g_*=106.75$, $\alpha_{\rm SM}=0.398$ (see below), $T_0=140$\,GeV and $V(0)^{1/4}=104$\,GeV one finds $\rho_R(T_0)/V(0)=115$, i.e.\ $\lambda_{\min}=1.1\times10^{-3}$ against $\lambda=0.129$. (With a single relativistic degree of freedom, $g_*=1$, the Standard Model would be marginal, $\rho_R(T_0)/V(0)=1.07$, and all the bounds below would be artificially strong, so the physical value of $g_*$ is essential here.)
This can be generalized to constrain the couplings of BSM scenarios. More specifically, the finite-temperature potential can be expressed as
\begin{equation}
    \label{eq:Finite_T_potential}
    V_{\text{eff}}(\phi,T) = V(\phi) + \frac{1}{2}\alpha T^{2} \phi^{2} + \cdots,
\end{equation}
where the $\cdots$ include higher-order terms in $\phi$, and $\alpha$ is the standard one-loop thermal-mass coefficient~\cite{Carrington:1991hz,Arnold:1992rz}. The condition for the symmetry-restoration temperature is
\begin{equation}
    \label{eq:Tc_cond}
    \frac{\partial^{2}V_{\text{eff}}(\phi)}{\partial\phi^{2}}\Bigg|_{\phi=0} = 0 = V''(0) + \alpha T_{0}^{2}~~\implies~~ T^{2}_{0} = \frac{-V''(0)}{\alpha},
\end{equation}
which when inserted into $\sigma T_0^4\geq cV(0)$ yields the following constraint
\begin{equation}
    \label{eq:FL_FOPT_bound}
    \frac{-V''(0)}{\sqrt{V(0)}} \geq \alpha\sqrt{\frac{c}{\sigma}}\,.
\end{equation}
Here $T_0$ is the spinodal temperature of the symmetric phase rather than the critical temperature $T_{\rm crit}\geq T_0$ of a first-order transition. Imposing the condition at $T_0$ makes it more restrictive than necessary, so that its failure is not an exclusion. A complete treatment would use the daisy-resummed thermal potential and the nucleation temperature, which we do not compute. What we quote below is therefore a diagnostic, whose boundary happens to lie within a few percent, in $C_H/\Lambda^2$, of the tree-level condition $\mu^2\geq0$ discussed below. At the boundary of the diagnostic found below, $T_0=33$\,GeV while $T_{\rm crit}=77$\,GeV and $\rho_R(T_{\rm crit})/V(0)=28$, and even where no spinodal exists ($\mu^2<0$, e.g.\ $T_{\rm crit}=73$\,GeV and $\rho_R(T_{\rm crit})/V(0)=26$ at $C_H/\Lambda^2=-2.9$\,TeV$^{-2}$) a transition completing by nucleation not far below $T_{\rm crit}$ would satisfy the requirement.

\begin{figure}[t]
    \centering
    \includegraphics[width=\textwidth]{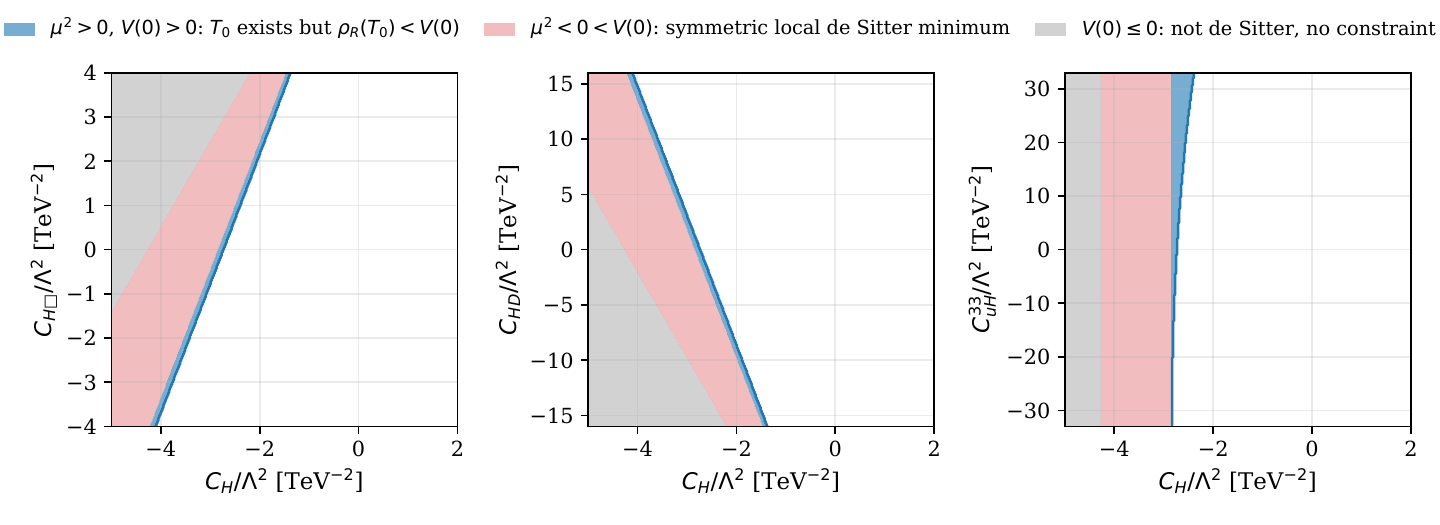}
    \caption{The three regions of the tree-level SMEFT potential in the planes spanned by $C_H$ and $C_{H\Box}$, $C_{HD}$ and $C^{33}_{uH}$ (all other coefficients set to zero): for $V(0)\leq0$ (gray) the symmetric point lies at or below the electroweak vacuum and is not a de Sitter state, so that the argument does not apply. For $\mu^2<0$ and $V(0)>0$ (red) the symmetric point is a local de Sitter minimum with massless charged particles, which the FL bound forbids~\cite{Montero:2021otb,Lee:2021cor} if it is required in every de Sitter vacuum of the theory, and which on a purely cosmological reading is excluded unless the transition completes by tunneling while radiation still dominates. In the thin blue stripe the symmetry is restored at $T_0>0$ but $\rho_R(T_0)<V(0)$, eq.~\eqref{eq:smeft_master} with $g_*=106.75$ and $c=1$, i.e.\ the sufficient condition fails there, which is a diagnostic and not an exclusion. The ranges are restricted to the region in which the dimension-six description is meaningful, $2|C_{H\Box}|v^2/\Lambda^2\leq0.5$, $|C_{HD}|v^2/2\Lambda^2\leq0.5$ and $|C^{33}_{uH}|v^2/2\Lambda^2\leq1$ (i.e.\ $0.5\leq Z\leq1.5$ and $|y_t-y_t^{\rm SM}|\leq1$). Within these ranges the boundary of the stripe is essentially the diagnostic for $C_H$, which scales with the wave-function factor of eq.~\eqref{eq:Zfactor}, $C_H/\Lambda^2\gtrsim-2.7\,Z$\,TeV$^{-2}$, and depends only mildly on $y_t$ ($-2.8$ to $-2.4$\,TeV$^{-2}$ over the range shown).}
    \label{fig:SMEFT_new}
\end{figure}

Let us now consider the SMEFT. In the so-called Warsaw basis~\cite{Grzadkowski:2010es}, the dimension-six operators that contribute to the tree-level potential are
\begin{align}
    \mathcal{L}^{\text{SMEFT}}_{\text{dim-6}} & \supset \frac{C_{H}}{\Lambda^{2}}(H^{\dagger}H)^{3} + \frac{C_{H\Box}}{\Lambda^{2}}(H^{\dagger}H)\Box (H^{\dagger}H)+ \frac{C_{HD}}{\Lambda^{2}} (H^{\dagger}D^{\mu}H)^{*}(H^{\dagger}D_{\mu}H)\nonumber\\
    & + \Big[\frac{C^{ij}_{uH}}{\Lambda^{2}}(H^{\dagger}H)(\overline{q}_{i} u_{j}\tilde{H})+{\rm h.c.}\Big], \label{eq:dim_6_SMEFT}
\end{align}
with the SM Yukawa coupling defined as $\mathcal L_{\rm SM}\supset-y_t\,\bar q_3\tilde Ht_R+{\rm h.c.}$ Other operators (e.g.\ $\mathcal O_{HW},\mathcal O_{HB},\mathcal O_{HWB}$) affect the thermal mass only through the gauge-boson masses at loop level and are neglected. The first operator modifies the tree-level potential directly, whereas the second and third modify the kinetic term and field renormalization, and the last modifies the mass of the corresponding up-type quark and contributes to the thermal potential at loop level. Of the operators $\mathcal O^{ij}_{uH}$ only $C^{33}_{uH}$, which modifies the top Yukawa coupling, is relevant, and we neglect all the others. To illustrate how the bounds are obtained, we consider the $C_{H}$ operator first while setting the remaining ones to zero. Setting $H = \begin{pmatrix} 0 & h/\sqrt{2}\end{pmatrix}^{T}$, the tree-level potential reads
\begin{equation}
    \label{eq:V_CH}
    V(h) = V(0) - \frac{1}{2}\mu^{2}h^{2} + \frac{1}{4}\lambda h^{4} - \frac{C_{H}}{8\Lambda^{2}}h^{6},
\end{equation}
where the constant $V(0)$ is fixed by requiring $V(v)=0$. Imposing the VEV and mass conditions
\begin{equation}
    \label{eq:VEV_m}
    V'(h)\Big|_{h=v} = 0, \hspace{5mm} V''(h)\Big|_{h=v} = m_{h}^{2},
\end{equation}
we obtain
\begin{equation}
\begin{rcases}
 \mu^{2} = \frac{1}{2}m_{h}^{2} +\frac{3C_{H}}{4\Lambda^{2}}v^{4}\\
\lambda = \frac{m_{h}^{2}}{2v^{2}} +\frac{3C_{H}}{2\Lambda^{2}}v^{2}
\end{rcases}
\quad \Longrightarrow \quad
V(0) = \frac{v^{2}}{8}\Big(m_{h}^{2} +\frac{C_{H}}{\Lambda^{2}}v^{4}\Big).
\end{equation}
Thus, the condition in eq.~\eqref{eq:FL_FOPT_bound} translates into the following bound on $C_{H}$
\begin{equation}
    \label{eq:CH_bound}
    \frac{(2m_{h}^{2}+\frac{3C_{H}}{\Lambda^{2}}v^{4})^{2}}{2v^{2}(m_{h}^{2}+\frac{C_{H}}{\Lambda^{2}}v^{4})} \geq \frac{c\,\alpha^2}{\sigma},
\end{equation}
where $\alpha=\frac{2m_W^2+m_Z^2}{4v^2}+\frac{\lambda}{2}+\frac{y_t^2}{4}$ reduces to $\alpha_{\rm SM}=\frac{1}{4v^{2}}(2m_{W}^{2}+m_{Z}^{2} + m_{h}^{2} + 2m_{t}^{2})=0.398$ in the Standard Model, but depends on the Wilson coefficients through $\lambda$ and $y_t$. Squaring has lost the sign of $\mu^2$: eq.~\eqref{eq:CH_bound} is to be read together with $\alpha>0$ and $\mu^2>0$, without which no spinodal temperature exists, and with $V(0)>0$, without which the symmetric point is not a de Sitter state. At $C_H/\Lambda^2=-3$\,TeV$^{-2}$, for instance, the squared inequality is satisfied although $\mu^2=-432$\,GeV$^2$. The unsquared form, eq.~\eqref{eq:smeft_master}, and Figure~\ref{fig:SMEFT_new} keep the distinction.

The operators $\mathcal O_{H\Box}$ and $\mathcal O_{HD}$ do not change the tree-level potential but rescale the kinetic term of $h$ at the minimum, $\tfrac12 Z\,(\partial h)^2$ with
\begin{equation}\label{eq:Zfactor}
    Z=1-\frac{2C_{H\Box}v^2}{\Lambda^2}+\frac{C_{HD}v^2}{2\Lambda^2}\,,
\end{equation}
so that the physical Higgs mass is $m_h^2=V''(v)/Z$ while $v=246$\,GeV remains the vacuum expectation value of the field that appears in the covariant derivative (it is fixed by $m_W$). Repeating the steps above with all four operators switched on (Appendix~\ref{app:smeft}) one finds
\begin{align}
    \mu^2&=\frac{m_h^2Z}{2}+\frac{3C_Hv^4}{4\Lambda^2}\,,\qquad
    \lambda=\frac{m_h^2Z}{2v^2}+\frac{3C_Hv^2}{2\Lambda^2}\,,\qquad
    V(0)=\frac{v^2}{8}\Big(m_h^2Z+\frac{C_Hv^4}{\Lambda^2}\Big)\,,\label{eq:smeft_params}\\
    y_t&=\frac{\sqrt2\,m_t}{v}+\frac{C^{33}_{uH}v^2}{2\Lambda^2}\,,\label{eq:yt_smeft}
\end{align}
and the condition~\eqref{eq:FL_FOPT_bound} becomes
\begin{equation}\label{eq:smeft_master}
\begin{aligned}
    \frac{\sqrt2\,m_h}{v}\;\frac{Z+\tfrac32\,\hat c_H}{\sqrt{Z+\hat c_H}}\;&\geq\;\sqrt{\frac{c}{\sigma}}\left[\frac{2m_W^2+m_Z^2}{4v^2}+\frac{m_h^2Z}{4v^2}+\frac{3C_Hv^2}{4\Lambda^2}+\frac{1}{4}\Big(\frac{\sqrt2\,m_t}{v}+\frac{C^{33}_{uH}v^2}{2\Lambda^2}\Big)^{2}\right],\\
    \hat c_H&\equiv\frac{C_Hv^4}{\Lambda^2m_h^2}\,.
\end{aligned}
\end{equation}
For a single operator this reduces to $(\sqrt2m_h/v)\sqrt Z\geq\alpha\sqrt{c/\sigma}$ for $\mathcal O_{H\Box}$ and $\mathcal O_{HD}$ (identifying $v$ with the vacuum expectation value of the canonically normalized field instead would change this at first order in the Wilson coefficient), and to a quadratic condition on $C^{33}_{uH}$ for $\mathcal O_{uH}$. The sign of the linear term in $C^{33}_{uH}$ is fixed by the conventions stated below eq.~\eqref{eq:dim_6_SMEFT}. We keep the exact tree-level dependence on $C_H$ rather than linearizing in $C_H/\Lambda^2$: the operator competes with the small quartic coupling $\lambda=0.13$, so that $\hat c_H$ is of order one where the bound lies although the expansion parameter of the effective theory, $C_Hv^2/\Lambda^2\simeq0.17$ at the bound, is small. This is the standard treatment of the $|H|^6$ operator in studies of the electroweak transition~\cite{Grojean:2004xy}. Contributions of other operators to the input relations at $\mathcal O(\Lambda^{-2})$ (e.g.\ to $m_Z$ through $C_{HD}$) are neglected. The resulting bounds are collected in Table~\ref{tab:1} and Figure~\ref{fig:SMEFT_new}.

\begin{table}[t!]
\centering
\footnotesize
\setlength{\tabcolsep}{4pt}
\resizebox{\textwidth}{!}{%
\begin{adjustbox}{max width=\linewidth}\begin{tabular}{lcc}
\toprule
Operator & this work, spinodal diagnostic ($g_*=106.75$, $c=1$) [TeV$^{-2}$] & collider / individual fit [TeV$^{-2}$] \\
\midrule
$\mathcal{O}_{H}$ & $C_{H}/\Lambda^2 \geq -2.73$\quad($-2.81$ for $c=0.1$, $-2.45$ for $c=10$) & $-10.9\lesssim C_H/\Lambda^2\lesssim 3.6$~\cite{CMS:2026nuu} \\
$\mathcal{O}_{H\Box}$ & $C_{H\Box}/\Lambda^2 \leq 8.2$ & $| C_{H\Box}/\Lambda^2|\lesssim 0.69$~\cite{deBlas:2025xhe} \\
$\mathcal{O}_{HD}$ & $C_{HD}/\Lambda^2 \geq -32.8$ & $| C_{HD}/\Lambda^2|\lesssim 0.024$~\cite{deBlas:2025xhe} \\
$\mathcal{O}_{uH}^{33}$ & $-166\leq C_{uH}^{33}/\Lambda^2 \leq 101$ & $|C_{uH}^{33}/\Lambda^2| \lesssim 1.56$~\cite{deBlas:2025xhe} \\
\bottomrule
\end{tabular}\end{adjustbox}}
\caption{Ranges of the dimension-six SMEFT operators that modify the Higgs potential and its thermal mass at $\mathcal O(\Lambda^{-2})$, in which the sufficient condition for a radiation-dominated electroweak transition, eq.~\eqref{eq:smeft_master}, is satisfied (its failure is a diagnostic, not an exclusion, Section~\ref{sec:SMEFT}), with one operator switched on at a time, compared with the ATLAS+CMS combination of Higgs-pair searches~\cite{CMS:2026nuu} (whose result $-0.71<\kappa_\lambda<6.1$ is converted with $\kappa_\lambda=1-2C_Hv^4/(m_h^2\Lambda^2)$) and with the one-operator-at-a-time fits of~\cite{deBlas:2025xhe}, quoted there as $95\%$ limits on the effective scale $\Lambda/\sqrt{|C|}$ ($1.2$, $6.5$ and $0.8$\,TeV) at a matching scale of $3$\,TeV with a $U(2)^5$ flavor symmetry, and converted here to $C/\Lambda^2$. Both comparisons are individual-coefficient limits rather than simultaneous global-fit intervals. Only the bound on $C_H$ lies within the range of validity of the effective description: at the other bounds the corrections they induce in the Higgs kinetic term or in the top Yukawa coupling are of order one.}
\label{tab:1}
\end{table}

With the physical number of degrees of freedom the resulting constraints are modest. Only the diagnostic on $C_H$ is meaningful: the sufficient condition fails for $C_H/\Lambda^2<-2.73$\,TeV$^{-2}$, which is only slightly above the requirement that the origin does not become a local minimum of the tree-level potential while remaining a de Sitter point, $\mu^2\geq0$, i.e.\ $C_H/\Lambda^2\geq-\tfrac23m_h^2/v^4=-2.84$\,TeV$^{-2}$~\cite{Montero:2019ekk,Montero:2021otb,Lee:2021cor} ($T_0\to0$ as $\mu^2\to0$, and the dependence on $c$ is mild, $-2.81$ to $-2.45$ for $c=0.1$--$10$). What does follow from the FL bound is the second condition, $\mu^2\geq0$, under the premise that the bound holds in every de Sitter vacuum of the theory~\cite{Montero:2021otb,Lee:2021cor}: a symmetric local minimum with $V(0)>0$ would be a de Sitter vacuum with massless charged particles. Within the dimension-six potential such a minimum exists precisely in the interval $-4.26\,{\rm TeV}^{-2}<C_H/\Lambda^2<-2.84\,{\rm TeV}^{-2}$ (at $Z=1$, the two ends are $V(0)=0$ and $\mu^2=0$, and the interval scales with $Z$). Below it the origin has $V(0)\leq0$ and is not a de Sitter point, so the argument does not apply there. At the upper end the quartic is negative, $\lambda=-m_h^2/2v^2$, so that the origin is a quartic maximum and not a minimum, and the boundary value itself is allowed. On the branch $V(0)>0$, where the symmetric stationary point has positive energy, the FL statement is therefore $\mu^2\geq0$, i.e.\ $C_H/\Lambda^2\geq-2.84$\,TeV$^{-2}$. The radiation-domination argument does not sharpen it, and the position of that boundary is itself sensitive to the truncation of the effective theory: a dimension-eight operator $C_8(H^\dagger H)^4/\Lambda^4$ at the scale $\Lambda=593$\,GeV implied by $C_H=-1$ at that boundary shifts $\mu^2$ by $C_8v^6/\Lambda^4=\pm1.8\times10^{3}$\,GeV$^2$ and the boundary by $-4C_8v^2/(3\Lambda^4)$, i.e.\ by $\mp0.65$\,TeV$^{-2}$ for $C_8=\pm1$. We quote this $23\%$ shift as an illustration of the truncation sensitivity at $|C_8|=1$, not as a defined or complete uncertainty of the effective-theory expansion. The bounds on the other three operators lie outside the range of validity of the dimension-six truncation and are quoted in Table~\ref{tab:1} only for completeness: the values $C_{H\Box}/\Lambda^2\leq8.2$\,TeV$^{-2}$ and $C_{HD}/\Lambda^2\geq-32.8$\,TeV$^{-2}$ are the points at which the wave-function factor~\eqref{eq:Zfactor} has fallen to $Z\simeq0.006$, i.e.\ at which the kinetic term of the Higgs nearly vanishes, and the interval $-166\leq C^{33}_{uH}/\Lambda^2\leq101$\,TeV$^{-2}$ corresponds to a top Yukawa coupling $|y_t|\simeq4$. They are all far weaker than the collider and electroweak-precision constraints. Figure~\ref{fig:SMEFT_new} shows the three regions of the potential, and the stripe in which the sufficient condition fails, in the planes spanned by $C_H$ and each of the other coefficients, obtained from eq.~\eqref{eq:smeft_master} with two operators switched on.

Regarding a strong first-order EWPT, which in the SMEFT requires a sizable positive $|H|^6$ term, i.e.\ $C_H<0$ in our conventions, with $\Lambda/\sqrt{|C_H|}\simeq0.5$--$0.85$\,TeV~\cite{Grojean:2004xy,Hashino:2022ghd}, the absence of a symmetric local minimum, $C_H/\Lambda^2\geq-2.84$\,TeV$^{-2}$, corresponds to $\Lambda/\sqrt{|C_H|}\geq0.59$\,TeV, at the lower edge of that window. The illustrative $23\%$ dimension-eight sensitivity noted above applies to that edge. The radiation-domination diagnostic adds nothing to it, and a sizable part of the window survives the FL condition. There is thus no established tension between the FL argument and a strong first-order electroweak phase transition induced by dimension-six operators, contrary to the conclusion that one would draw with $g_*=1$. The constraint in this subsection uses only the condition of radiation domination at $T_0$~\cite{Venken:2023hfa}, whereas the absence of a symmetric de Sitter minimum rests on the FL bound in that vacuum. The thermal bound~\eqref{eq:TFL_final} itself adds nothing to it, because in the symmetric phase that bound is violated by construction, all charged particles having $m\lesssim gT\ll\lc T$. That is exactly why the phase transition has to occur in radiation domination.

\section[Inflation]{Inflation}
\label{sec:inflation}\label{sec:slowroll}

So far we have applied the thermal bound to the present cosmological epoch. Since the bound scales as $V^{1/4}$, it is natural to ask what it implies for the early universe, where for slow-roll inflation the vacuum energy would be very large and our conjecture applies.

During slow roll the vacuum energy is constant to slow-roll accuracy and the conjecture applies with $H=H_I$, $V_I\simeq3\MP^2H_I^2$. In cold inflation the only bath is the Gibbons--Hawking radiation~\cite{Gibbons:1977mu} at $T_{\rm GH}=H_I/2\pi$, which lies below $\Tc$ for every mass $m<\pi\sqrt6\,gq\MP$. The zero-temperature bound~\eqref{eq:FL_log} then applies, with $\lS=\ln[(\sqrt6/2\pi^3)(gq)^3\MP/H_I]$ ($5.6$ for $gq=e$ and $H_I=10^{13}$\,GeV). A thermal bath during inflation, as in warm inflation~\cite{Berera:1995ie,Bartrum:2013fia,Bastero-Gil:2016qru,Kamali:2023lzq}, adds the thermal bound~\eqref{eq:TFL_final} with
\begin{equation}\label{eq:Tmax_inflation}
    \Tmax(V_I)=\Big(\frac{cV_I}{\sigma}\Big)^{1/4}=0.54\,c^{1/4}\sqrt{\MP H_I}=3.4\,c^{1/4}\sqrt{\frac{\MP}{H_I}}\;T_{\rm GH}\qquad(g_*=106.75)\,,
\end{equation}
but such a bath must be a small fraction of the energy budget: with $\rho_R=cV_I$ the expansion rate is $H^2=(1+c)H_I^2$ and $c=1$ is not slow roll at all ($\epsilon_H=2c/(1+c)=1$), while the radiation sustained by dissipation in warm inflation has $\rho_R/V\simeq\epsilon_HQ/2(1+Q)\ll1$ in terms of the dissipation ratio $Q$. Since $\Tmax\propto c^{1/4}$, the thermal bound weakens as $c$ decreases and falls below the zero-temperature bound for small $c$: at $c=0.034$ for  a Higgs displaced to $\MP$ with $g_*=106.75$, at $c=0.007$ for the same displacement with $g_*=23.25$, and below $c=10^{-4}$ for an electron at its electroweak mass, for which only the photons and the neutrinos are relativistic. The zero-temperature statement is therefore the robust one during inflation, and we quote the thermal numbers for $c=1$, the strongest possible constraint.

For the electron in the electroweak vacuum the zero-temperature bound is violated for $5\times10^{-27}\,{\rm GeV}<H_I<2.7\times10^{15}$\,GeV: the lower edge is $H_I\simeq\pi m_e^2/(\lS\sqrt6\,e\MP)$, i.e.\ $V_I^{1/4}\simeq0.14$\,MeV, and the upper edge is the scale $(\sqrt6/2\pi^3)e^3\MP$ at which $\lS$ vanishes and the semiclassical criterion~\eqref{eq:pair_criterion} becomes empty (the bound $m^2\geq(\lS/\pi)\sqrt6\,e\MP H_I$ has two roots in $H_I$ because $H_I\lS(H_I)$ vanishes at both ends). For the standard vacuum tensor spectrum in Einstein gravity, $P_t=2H_*^2/(\pi^2\MP^2)=A_sr$, the observational limit on the tensor-to-scalar ratio, $r<0.036$~\cite{BICEPKeck:2022mhb}, gives $H_*<4.7\times10^{13}$\,GeV at the pivot scale, and $\dot H\leq0$ carries that ceiling forward to the subsequent inflationary epochs (it says nothing about earlier ones, where $H$ may have been larger). Successful nucleosynthesis requires, on the other hand, $H_I\geq H(T_{\rm BBN})=7\times10^{-24}$\,GeV for $T_{\rm BBN}\simeq4$\,MeV. The whole interval $7\times10^{-24}\,{\rm GeV}\leq H_*<4.7\times10^{13}$\,GeV therefore lies inside the excluded range, so that slow-roll inflation compatible with the CMB limit on tensor modes and with nucleosynthesis is excluded within the domain of the conjecture (Appendix~\ref{sec:validity}).

There are analogous bounds from all the other electrically charged Standard Model particles, whereas neutrinos charged under an unbroken $B-L$ give none: for $g_{B-L}\lesssim10^{-24}$ the logarithm $\lS$ is negative at every $H_I\geq H(T_{\rm BBN})$, as it is today (Section~\ref{sec:neutrinos}), and $g_{B-L}$ lies far below $(gq)_{\min}$. It is clear that if one wishes to have a period of slow-roll inflation compatible with our bound a drastic modification of the Standard Model sector is required.

The most natural modification would be to have the Higgs play the role of the inflaton. This is minimal in that one does not need to introduce any new fields: not even the inflaton is a new field. All that is needed is to modify the Higgs potential at large values of the radial scalar $\phi$. One needs to ensure that wherever the potential is positive and either slow-roll or in a minimum, $\phi$ is sufficiently large such that all charged particles are sufficiently heavy and obey our bound.

Unfortunately vanilla Higgs inflation \cite{Bezrukov:2007ep}, which we briefly review in Appendix~\ref{sec:higgsreview}, violates our bound. As was already pointed out in the original FL paper \cite{Montero:2019ekk}, for a fermion with Yukawa coupling $Y$ to the Higgs-inflaton, $m=Y\phi/\sqrt2$ and $V_J=\lambda\phi^4/4$ in the Jordan frame, so that
\begin{equation}\label{eq:Y4lambda}
    \frac{m^4}{V}=\frac{Y^4}{\lambda}\,,
\end{equation}
independently of the field value and of the non-minimal coupling $\xi$. The ratio is invariant under the conformal transformation to the Einstein frame, $m_E=m_J/\sqrt f$ and $V_E=V_J/f^2$, and on the plateau, $\xi\phi^2\gg\MP^2$, it gives $V_E\to\lambda\MP^4/4\xi^2$ and $m_E\to Y\MP/\sqrt{2\xi}$, the masses of a Higgs displaced to $\MP/\sqrt\xi$. Even the zero-temperature bound~\eqref{eq:FLoriginal}, $m^4\geq2Q^2V$ with $Q=2e/3$ for the top quark, then requires $Y\geq(2Q^2\lambda)^{1/4}\simeq0.17\text{--}0.30$ for $\lambda\simeq0.01$--$0.1$: it is satisfied by the top quark only, and violated by the electron by twenty orders of magnitude in $m^4/V$. Take $\xi=10^4$, $\lambda=0.038$ and $y_t=0.5$ at the inflationary scale. Then $H_I=1.4\times10^{13}$\,GeV and $m_t=8.6\times10^{15}$\,GeV, against the $5.2\times10^{15}$\,GeV that eq.~\eqref{eq:FL_log} requires once the three colors of the top quark are counted in the current, so the top quark passes. It would not pass the formal $c=1$ thermal bound, which asks for $1.2\times10^{16}$\,GeV. The electron, at $5\times10^{10}$\,GeV, fails by five orders of magnitude. The observation of~\cite{Montero:2019ekk} that in ``vanilla'' Higgs inflation the light fermions remain too light therefore holds a fortiori for the thermal bound.

We then see that the bound is violated, but not parametrically, since $m^4/V$ does not grow with the field value. The violation is a result of the numerically small Yukawa couplings. If one is fond of adding epicycles one may further try to engineer a scenario where the Yukawa couplings during inflation differ from those today and satisfy our bound in this way.

A more natural resolution might seem to come from the observation  that the canonically normalized inflaton is trans-Planckian during the inflationary era (it reaches $5.5\MP$ for $\xi=10^4$, whereas $\phi_I\simeq0.09\MP$, Table~\ref{Tab:2}). One then expects that quantum gravity corrections to the Higgs potential will become relevant. One may wonder to what extent one can then trust an EFT analysis of inflation, but let us assume that these corrections can be treated within the EFT. Such corrections cannot remove the violation, however, whatever their sign. In the Einstein frame the electron mass $m_e=y_e\phi/\sqrt{2f(\phi)}$ never exceeds $y_e\MP/\sqrt{2\xi}$, i.e.\ $5\times10^{10}$\,GeV for $\xi=10^4$, independently of the potential, and eq.~\eqref{eq:FL_log} then requires $H_I\lesssim1.5\times10^{2}$\,GeV, eleven orders of magnitude below the CMB-normalized plateau, $H_I\simeq1.4\times10^{13}$\,GeV. For any $\xi$, the bound requires the effective quartic coupling $\lambda_{\rm eff}=4V_J/\phi^4$ at the field values of inflation to satisfy $\lambda_{\rm eff}\lesssim\pi^2y_e^4/(2e^2\lS^2)\simeq10^{-22}$ for $H_I=10^{13}$\,GeV, the logarithmic form of the condition of~\cite{Lee:2021cor}. Since the inflaton has to roll into the electroweak vacuum, the potential must then be modified already at $\phi\sim10^{11}$\,GeV for $\xi=10^4$, far below the Planck scale. This differs from the weak gravity conjecture, where the sign of the higher-derivative corrections to Einstein--Maxwell theory suffices for small charged black holes to satisfy it~\cite{Arkani-Hamed:2021ajd}.

Alternatively, one may also consider scenarios where the Higgs plays the role of the inflaton in scenarios other than vanilla Higgs inflation. For example, one might have a metastable minimum for the Higgs at large $V$ and $\phi$ before tunneling to our present epoch. The zero-temperature FL bound was already considered in this context in~\cite{Lee:2021cor}. It is straightforward to apply our bound to whatever scenario one likes analogously to the discussion in this section. 

Instead of, or in conjunction with, Higgs inflation one may consider a scenario where the inflaton is coupled directly or indirectly to the gauge fields in the theory. One can then engineer a scenario where all gauge fields are either Higgsed or confined with a scale above the inflationary Hubble scale or their gauge coupling strength is so weak that our bound does not apply. One might worry that such scenarios would be rather fine-tuned and contrived. However, in~\cite{Nakamura:2026qvp} a fairly clean mechanism to Higgs all gauge fields during inflation within the MSSM was proposed. Compellingly, taking this route to satisfy FL seems to also provide a channel for baryogenesis.

\section{Discussion}
\label{sec:Discussion}

We have extended the Festina Lente argument to de Sitter backgrounds that contain an equilibrium thermal bath. Our main result is a discharge constraint from conduction by the thermal population of the charged species. It is stronger than the photon-fusion constraint in their common domain of validity, without requiring a comparison with every thermal production channel: once the discharge criterion is formulated as the loss of an order-one fraction of the black-hole charge $\QN\sim10^{61}/gq$ within a Hubble time, eq.~\eqref{eq:criterion}, the bath must not contain more than $\sim\sqrt6\MP H^2/(gq)$ charge carriers per unit volume, which for a Boltzmann-suppressed species is the condition $m\geq\lc T$ of eq.~\eqref{eq:TFL}. For the maximal bath allowed by vacuum domination this is a bound of order $V^{1/4}$ times a logarithm of order $10$--$70$, which depends on the charge and gauge coupling only logarithmically and switches off only for $gq\lesssim10^{-30}$ at the present vacuum energy (between $3\times10^{-31}$ and $3\times10^{-29}$ within the sensitivity band). The same discharge criterion rescales the zero-temperature FL bound by the factor $\lS/\pi$ in $m^2$, eq.~\eqref{eq:FL_log}, which sharpens it for $gq\gtrsim7\times10^{-20}$, and shows that semiclassical pair creation cannot discharge a Nariai black hole within a Hubble time for $gq<2\times10^{-20}$ today. Both statements follow from evaluating explicitly the discharge time whose parametric form was given in~\cite{Montero:2019ekk} (their Section~4.1.1), and from its extension to the equilibrium conduction current. We have also shown that, for the thermal population, violating the discharge criterion means that the carriers between the horizons screen the field. In homogeneous models of the throat, the conversion of the field energy into those accelerated carriers starts a contraction that reaches a curvature singularity in a time of order $H^{-1}$, so no assumption about their thermalization enters the criterion. The Hubble-time threshold expresses the requirement that this conversion be complete before the inhomogeneous modes of the throat can take over (Section~\ref{sec:criterion}). That the singularity is then naked is the assumption that the argument shares with~\cite{Montero:2019ekk}. The order-one choices in the criterion shift the logarithms by about $\pm3$, which we carry as a sensitivity band.

Phenomenologically, and for the formal $c=1$ benchmark\footnote{Let us stress one final time that in the absence of a top-down derivation of the value of $c$, the largest value of $c$ where our bound could plausibly hold is $c=1$, leading to the strongest possible constraint. If the actual value of $c$ turns out to be smaller, the bound will be correspondingly weakened.} with the minimal bath, the thermal bound requires every particle with a charge $10^{-26}\lesssim gq\lesssim e$ under a $U(1)$ whose gauge boson is lighter than $H_\Lambda$ to be heavier than $0.03$--$0.2$\,eV (Table~\ref{tab:mmin}). For an unbroken gauged $B-L$ with Dirac neutrinos this means $m_{\nu,{\rm lightest}}\gtrsim10$--$43$\,meV over the whole range $10^{-29}\lesssim g_{B-L}\lesssim10^{-24}$, which the DESI\,DR2 limit $m_{\nu,{\rm lightest}}<23$\,meV excludes for $g_{B-L}\gtrsim7\times10^{-28}$ and which is within reach of the next generation of cosmological surveys. A neutrino millicharge must satisfy $\epsilon_\nu\lesssim2\times10^{-27}$ if that limit holds, and ultralight millicharged particles with $\epsilon\gtrsim10^{-28}$ are excluded below $10^{-2}$--$0.2$\,eV. These exclusions weaken quickly for a cooler bath: for $c=0.1$, at the central normalization of the current, the DESI exclusion of an unbroken $B-L$ moves above the laboratory limit on $g_{B-L}$ (Section~\ref{sec:neutrinos}). The constraints on the SMEFT from the electroweak phase transition, on the other hand, are modest once the physical number of relativistic degrees of freedom is used: the radiation-domination requirement gives only a sufficient-condition diagnostic that does not go beyond the absence of a symmetric de Sitter minimum, $C_H/\Lambda^2\geq-2.8$\,TeV$^{-2}$ on the branch where the symmetric point has positive energy (a boundary whose position carries an illustrative $23\%$ dimension-eight sensitivity), and there is no established tension with a strong first-order transition. During standard slow-roll inflation the conjecture is violated by the electron for any $H_I$ between $5\times10^{-27}$\,GeV and $2.7\times10^{15}$\,GeV. That interval contains the whole range of Hubble rates compatible with the CMB limit on tensor modes (for the standard vacuum tensor spectrum) and with nucleosynthesis. We have seen that one may in principle engineer slow-roll inflationary scenarios compatible with our bound by coupling to the Standard Model sector, but the most vanilla version of Higgs inflation does not work and one is heavily constrained in what scenarios could in principle work.

Several caveats should be kept in mind, and they define the directions in which this work should be continued.
\begin{enumerate}
\item \emph{The FL argument itself.} The conclusion that rapid discharge leads to a violation of cosmic censorship rests on the analysis of~\cite{Montero:2019ekk}, which has been revisited in~\cite{Aalsma:2023mkz}, where decay channels that circumvent the bound without producing a naked singularity were identified. Our results inherit this uncertainty. We note, however, that the two analyses concern different regimes. In the rapid regime the discharge is a collective, classical process of the $\sim\QN$ carriers between the horizons, involving all angular momenta on the $S^2$. In the homogeneous models of Appendix~\ref{app:discharge} it contracts the throat through the loss of the tension of the field (Section~\ref{sec:criterion}). Those models establish the onset and the time scale of the contraction, but not its global outcome. The channels of~\cite{Aalsma:2023mkz} are, as we read them, the emission of individual spherically symmetric shells, with a probability set by the entropy difference between two static solutions, and they are unsuppressed when that difference is small. Whether the emission of many particles at once, which is our rapid regime, dominates instead is left open in~\cite{Aalsma:2023mkz}. What the FL argument assumes is that the rapid regime, when it is kinematically available, is the one that takes place, as its time scale suggests. A treatment of the discharge in the $dS_2\times S^2$ geometry with the backreaction of the plasma and with inhomogeneous perturbations would settle this point, and it would also fix the order-one constants inside the logarithms, which we have only been able to estimate ($\delta\lc\simeq\pm3$, Section~\ref{sec:TFLbound}). Until then the conjecture of Section~\ref{sec:criterion} is a conditional diagnostic, restricted to near-ultracold black holes, whose consequences we have worked out within that restriction. In support of the bound, we recall that it is satisfied by the charged spectrum of the Standard Model at the present vacuum energy~\cite{Montero:2019ekk}, that it is consistent with the throat region of warped anti-brane constructions in string theory~\cite{Montero:2021otb}, and that the Standard Model Higgs sector satisfies the finite-temperature version proposed in~\cite{Venken:2023hfa}. A direct check of the thermal bound~\eqref{eq:TFL_final} in a particular realization would be very valuable.
\item \emph{The hypothetical bath.} The bound tests the consistency of the EFT in a state, the maximal bath, that the universe is not in. This is the same logic by which the FL bound tests the theory with black holes that have never formed, but a reader who rejects swampland-type reasoning about hypothetical states will regard eq.~\eqref{eq:TFL_final} as applying only to universes that actually contain such a bath. The order-one parameter $c$ and the composition of the bath enter the bounds through $\Tmax\propto(c/g_*)^{1/4}$ and the logarithm, and the endpoint $(gq)_{\min}$ as $(g_*/c)^{3/4}$. The bath itself destabilizes the Nariai throat, within $0.77/H$ for $c=1$ and $3.3/H$ for $c=0.01$ (Appendix~\ref{app:discharge}), so the bounds for $c=1$ are a formal envelope. The controlled statements are those for $c\ll1$, where the bath is a small gravitational perturbation and the throat survives for several Hubble times.
\item \emph{Equilibrium and relics.} We have assumed the charged species to be in equilibrium with the bath. Non-equilibrium populations can be far more abundant: the relic neutrino background, with $n_\nu\simeq336\,{\rm cm}^{-3}$, would by the same argument discharge a $B-L$ Nariai black hole within a Hubble time for any $g_{B-L}\gtrsim3\times10^{-27}$, irrespective of the neutrino masses. We leave open whether the conjecture should be extended to such states. If it were, unbroken gauged $B-L$ with $m_{B-L}<H_\Lambda$ would be excluded in our universe for every $g_{B-L}$ above that threshold. Here we have deliberately restricted it to equilibrium states.
\item \emph{Inflation.} We have seen that slow-roll inflation generically violates our bound. Nevertheless, one can still engineer scenarios that obey our bound. On the one hand, it would be interesting to work these scenarios out in more detail and see if they lead to any specific observable predictions. On the other hand, one might worry that such scenarios could be rather contrived\footnote{Note, however, that very recently \cite{Nakamura:2026qvp} proposed an inflationary scenario where FL is satisfied via flat directions in the MSSM in a way which seems fairly clean.}. In this case, it is tempting to conclude that our bound is further evidence that slow-roll inflation is in tension with quantum gravity, in line with, e.g.,~\cite{Martin:2000xs,Bedroya:2019tba}. If so, it would be interesting to see if one can develop swampland bounds in a similar spirit to Festina Lente for alternative early universe scenarios.
\item \emph{Other applications.} The bound applies to any dark sector with an unbroken $U(1)$ and an ultralight gauge boson~\cite{Ackerman:2008kmp,Fabbrichesi:2020wbt,Montero:2022jrc}: dark-matter candidates charged under a massless dark photon must be heavier than $\lc\Tmax\simeq10^{-2}$--$0.2$\,eV for $gq\gtrsim10^{-28}$ in the $c=1$ benchmark, which excludes ultralight charged dark matter over that range, and models with a dark photon lighter than $H_\Lambda$ that kinetically mixes with the photon inherit the millicharge bounds of Section~\ref{sec:mCP}. A thermal version of the magnetic FL bound~\cite{Montero:2021otb} and of the axionic FL bound~\cite{Guidetti:2022xct} would also be worth deriving.
\end{enumerate}

\acknowledgments
FA thanks SOKN Engineering for their hospitality. VV is supported by World Premier International Research Center Initiative (WPI), MEXT, Japan.
NO is supported in part by the U.S. Department of Energy under Grant No.~DE-SC0023713. ZWW is partially supported by the National Natural Science Foundation of China (Grant No.~12475105, Grant No.~12622508).
{\it The use of AI disclosure}: Parts of the numerical cross-checks and the editing of the text were carried out with the help of AI assistants. All results were checked by the authors, who take full responsibility for the content.

\appendix

\section{The thermal Schwinger exponent}\label{app:instanton}
For completeness we re-derive the exponent of eq.~\eqref{eq:Scwhinger-T} in the worldline-instanton formalism~\cite{Affleck:1981bma,Dunne:2005sx,Brown:2015kgj,Gould:2017fve}. In Euclidean space with the time direction compactified on a circle of circumference $\beta=1/T$, the action of a closed worldline of a particle of mass $m$ and charge $gq$ in a constant electric field $E$ is $S=mL-gqE\,A$, where $L$ is the length of the loop and $A$ the area it encloses in the plane spanned by the Euclidean time and the field direction. Stationary loops are arcs of circles of radius $R=m/(gqE)$. For $\beta>2R$ the full circle fits inside the thermal cylinder and $S=2\pi mR-\pi gqER^2=\pi m^2/(gqE)$, the zero-temperature Schwinger exponent. For $\beta<2R$, i.e.\ $T>\Tc=gqE/2m$, the minimum-action loop is the lens formed by two arcs of radius $R$ whose common chord has length $\beta$~\cite{Brown:2015kgj}. The cusps are resolved by the photon exchange between the two arcs at scales of order $\alpha R$~\cite{Gould:2017fve}. With $\sin\theta=\beta/2R=\Tc/T$, each arc subtends the angle $2\theta$, so that
\begin{equation}
    L=4R\theta\,,\qquad A=2\left(R^2\theta-\tfrac{\beta}{2}R\cos\theta\right)=2R^2\theta-\beta R\cos\theta\,,
\end{equation}
and
\begin{equation}
    S_T=4mR\theta-gqE\left(2R^2\theta-\beta R\cos\theta\right)=\frac{2m^2}{gqE}\arcsin\frac{\Tc}{T}+\frac{m}{T}\sqrt{1-\frac{\Tc^2}{T^2}}\,,
\end{equation}
which is eq.~\eqref{eq:Scwhinger-T}. It is continuous at $T=\Tc$, where $\theta=\pi/2$ and $S_{\Tc}=\pi m^2/(gqE)$. For $T\gg\Tc$, writing $y=\Tc/T$ and using $(2m^2/gqE)\,y=m/T$,
\begin{equation}
    \begin{aligned}
    S_T&=\frac{m}T\left[\left(y+\frac{y^3}{6}+\ldots\right)\frac{1}y+1-\frac{y^2}{2}+\ldots\right]=\frac{m}T\left(2-\frac{y^2}{3}+\mathcal O(y^4)\right)\\
    &=\frac{2m}{T}\left(1-\frac{\Tc^2}{6T^2}+\ldots\right),
    \end{aligned}
\end{equation}
which is eq.~\eqref{eq:HighT_correct}. The leading term is the Boltzmann factor of a pair at rest, and the correction is $-(gqE)^2/(12m^2T^2)$ relative to $2$.

\section{Discharge times}\label{app:discharge}

\paragraph{Nariai geometry.} With $G=1/8\pi\MP^2$ and $\Lambda=3H^2$, the Reissner--Nordstr\"om--de Sitter metric function is $f(r)=1-2GM/r+GQ^2/(4\pi r^2)-H^2r^2$, where $Q$ is the charge in canonical units so that $E=Q/(4\pi r^2)$. Degenerate horizons, $f=f'=0$, occur on the curve $(Q\sqrt{G/4\pi}\,H,\,GMH)=(r\sqrt{1-3r^2},\,r-2r^3)$ with $r=r_{\rm hor}H$, shown in Figure~\ref{fig:BHdiagram}. The triply degenerate (ultracold) point~\cite{Romans:1991nq}, $f=f'=f''=0$, is at $r_{\rm hor}^2=1/6H^2$, where $GQ^2/(2\pi r_{\rm hor}^4)=6H^2$, i.e.\ $8\pi GE^2=6H^2$ and $E=\sqrt6\MP H$. This is the largest field on the Nariai branch~\cite{Montero:2019ekk} (along the degenerate-horizon curve $E^2=2\MP^2(r_{\rm hor}^{-2}-3H^2)$, which keeps growing toward smaller extremal black holes). The number of elementary charges follows from $Q=4\pi r_{\rm hor}^2E$: $\QN=Q/gq=(4\pi/\sqrt6)\MP/(gqH)$, eq.~\eqref{eq:QN}.

\paragraph{Kantowski--Sachs form and the fate of the field energy.} Between its horizons the near-Nariai black hole is the product space $dS_2\times S^2$ (at the ultracold point $\mathbb R^{1,1}\times S^2$)~\cite{Romans:1991nq,Bousso:1996au}, which in Kantowski--Sachs coordinates~\cite{Kantowski:1966te} reads $\dd s^2=-\dd t^2+a(t)^2\dd x^2+b(t)^2\dd\Omega^2$, with the electric field along $x$. For a homogeneous source with energy density $\rho$ and pressures $p_\parallel$ along $x$ and $p_\perp$ on the sphere, the Einstein equations with $\Lambda=3H^2$ are
\begin{align}
    \frac{\dot b^2}{b^2}+\frac{1}{b^2}+\frac{2\dot a\dot b}{ab}&=\Lambda+8\pi G\rho\,,\label{eq:KS_tt}\\
    \frac{2\ddot b}{b}+\frac{\dot b^2}{b^2}+\frac{1}{b^2}&=\Lambda-8\pi G\,p_\parallel\,,\label{eq:KS_xx}\\
    \frac{\ddot a}{a}+\frac{\ddot b}{b}+\frac{\dot a\dot b}{ab}&=\Lambda-8\pi G\,p_\perp\,.\label{eq:KS_thth}
\end{align}
The electric field has $\rho_E=E^2/2$, $p_\parallel=-\rho_E$ and $p_\perp=+\rho_E$. Static solutions, $b=b_0$, require from eq.~\eqref{eq:KS_xx} $1/b_0^2=\Lambda+8\pi G\rho_E$, and eq.~\eqref{eq:KS_thth} then gives $\ddot a/a=\Lambda-8\pi G\rho_E$, a $dS_2$ factor whose curvature vanishes when $8\pi G\rho_E=\Lambda$: this is the ultracold point, $E_N^2/2=3\MP^2H^2$ and $b_0^2=1/6H^2$, in agreement with eq.~\eqref{eq:EN}. Eq.~\eqref{eq:KS_xx} involves only the radius of the sphere and the pressure along the field lines, and it decides whether the throat starts to contract. If the field energy is converted, at fixed $\rho$, into a component with $p_\parallel>-\rho_E$, the right-hand side drops below $1/b_0^2$ and the sphere starts to contract,
\begin{equation}\label{eq:KS_collapse}
    \ddot b=-4\pi G\,b_0\,(p_\parallel+\rho_E)\,,\qquad
    \frac{\ddot b}{b_0}=\begin{cases}-3H^2&\text{counter-streaming carriers, }p_\parallel=\rho\,,\\ -2H^2&\text{radiation, }p_\parallel=\rho/3\,,\\ -\tfrac32H^2&\text{dust, }p_\parallel=0\,.\end{cases}
\end{equation}
The subsequent evolution follows from eqs.~\eqref{eq:KS_tt}--\eqref{eq:KS_thth} together with the conservation law $\dot\rho+A(\rho+p_\parallel)+2B(\rho+p_\perp)=0$ for each component ($A=\dot a/a$, $B=\dot b/b$): the field obeys $\rho_E\propto b^{-4}$ (flux conservation), and a component converted from it at $t=0$ with pressures $(w_\parallel,w_\perp)\rho$ obeys $\rho\propto a^{-(1+w_\parallel)}b^{-2(1+w_\perp)}$. Integrating numerically from the static ultracold throat (%
the integrations are stopped where $a$ or $b/b_0$ first reaches $10^{-6}$)\footnote{The quoted times are insensitive to this choice: moving the stopping surface from $10^{-3}$ to $10^{-6}$ changes them by less than $3\times10^{-4}H^{-1}$ and tightening the tolerances by two orders of magnitude leaves them unchanged to nine digits ($0.654894765\,H^{-1}$ for the streams). Up to the last point with $a>10^{-4}$ and $b/b_0>10^{-2}$ the Hamiltonian constraint~\eqref{eq:KS_tt} is violated by less than $10^{-10}$ of the sum of the absolute values of its terms, which is the computed norm  (the same bound holds with the largest single term in the denominator: the maxima are $5\times10^{-12}$ and $1.1\times10^{-11}$ for the complete stream conversion and $3.8\times10^{-11}$ and $9.0\times10^{-11}$ for the fastest continuous discharge). Measured against the much smaller combination $\Lambda+8\pi G\rho$, which the individual terms exceed by many orders of magnitude near the crunch, the same violation is at most $1.2\times10^{-6}$ for complete conversion and $8\times10^{-6}$ for the weakly charged and near-vacuum runs, and it falls by two orders of magnitude when the tolerances are tightened, i.e.\ it is a discretization error and not a drift of the solution. The only calculations that do not reach the stopping surface are the two slowest continuous discharges below, whose step size underflows at $(a,b/b_0)=(4.7\times10^{-5},1.6\times10^{-5})$ and $(4.3\times10^{-6},3.3\times10^{-5})$ respectively. Their terminal times agree to six digits with the runs stopped at $10^{-5}$, which do terminate on the surface, and we quote them on that basis rather than as completed integrations. The curvature invariant $R_{\mu\nu}R^{\mu\nu}=4\Lambda^2-2\Lambda\mathcal T/\MP^2+(\rho^2+p_\parallel^2+2p_\perp^2)/\MP^4$, with the trace $\mathcal T=-\rho+p_\parallel+2p_\perp$, grows by tens of orders of magnitude along these runs, as expected for a curvature singularity, although a finite numerical run is of course not a proof that one is reached.}, complete conversion of the field energy into counter-streaming carriers ($w_\parallel=1$, $w_\perp=0$) brings the sphere to $b=0$ after $t=0.65\,H^{-1}$, conversion into radiation ($w_\parallel=w_\perp=1/3$) after $0.75\,H^{-1}$ and into dust after $0.83\,H^{-1}$. If only half of the field energy is converted into radiation, the sphere contracts to $0.30\,b_0$ while the $dS_2$ direction collapses, $a\to0$ at $t=1.09\,H^{-1}$, with the same outcome: a curvature singularity that fills the whole throat. These times refer to the near-ultracold black holes that set the bound, $E\simeq\EN$. For a Nariai black hole with $E^2/\EN^2=\chi\ll1$ the same equations give $b=0$ after $1.6\,H^{-1}$ ($\chi=0.1$), $2.9\,H^{-1}$ ($10^{-2}$) and $5.6\,H^{-1}$ ($10^{-4}$), which is the time scale of the ordinary gravitational instability of the Nariai geometry~\cite{Bousso:1996au} rather than of the discharge. The collapse of the sphere is the crunch of~\cite{Montero:2019ekk}. Its onset follows from the loss of the tension of the field alone: the focusing equation $\dd(\dot b/a)/\dd t=-(\rho+p_\parallel)b/(2\MP^2a)$ makes $\dot b$ negative and keeps it so for any source obeying the null energy condition. The time scale, on the other hand, is that of the products. For a component with $p=(-1+\epsilon)\rho$ the crunch time grows without bound as $\epsilon\to0$ ($2.4\,H^{-1}$ for $\epsilon=0.01$, $2.8\,H^{-1}$ for $\epsilon=0.003$), so the times above are those of the three channels integrated here, which are the ones the discharge produces. Maxwell's equations in the same geometry reduce, by homogeneity, to $\dd(b^2E)/\dd t=-b^2J$, i.e.\ $\dot E=-J$ as long as the sphere is static (no magnetic field is generated at this level), and the energy of the field is transferred to the carriers at the rate $JE$ while nothing leaves the throat. Finally, the bath itself is not inert in the throat: a homogeneous radiation component $\rho_R=cV$ with $p_\parallel=\rho_R/3$ added to the static throat gives $\ddot b/b_0=-2cH^2$ at $t=0$ and, without any discharge, drives the throat to a crunch through its $dS_2$ direction ($a\to0$, with the sphere at $b/b_0=0.34$, $0.72$ and $0.90$ at that moment) after $0.77\,H^{-1}$ for $c=1$, $1.7\,H^{-1}$ for $c=0.1$ and $3.3\,H^{-1}$ for $c=0.01$. This is why the $c=1$ bounds are called a formal envelope in Section~\ref{sec:criterion}. The lifetime of the throat in the bath enters the criterion~\eqref{eq:criterion} through the number $\kappa$ of Hubble times available for the discharge. A continuous discharge can be followed with two cold species of mass $m$ and conserved number density $n=n_0/(a\,b^2/b_0^2)$, carrying the momenta $\pm p$ along the field, $\dot p=gqE-\dot a\,p/a$, with $\rho_{\rm c}=2n\sqrt{m^2+p^2}$, $p_\parallel=2np^2/\sqrt{m^2+p^2}$, $p_\perp=0$, the current $J=2gq\,np/\sqrt{m^2+p^2}$ and Maxwell's equation $\dd(b^2E)/\dd t=-b^2J$. These conserve the total energy, and they reverse the current when the carrier momentum reverses, which lags the reversal of the field: for $t_{\rm c}=0.05\,H^{-1}$ the field reverses at $0.0501\,H^{-1}$ and the current at $0.0997\,H^{-1}$. Starting from the static ultracold throat (no bath, carriers at rest, $2nm\ll\rho_E$) with $n_0$ chosen so that the current saturates at $J=\EN/t_{\rm c}$, i.e.\ so that the frozen-geometry discharge time is $t_{\rm c}$, the crunch occurs at $0.75$, $0.76$, $0.81$ and $0.95\,H^{-1}$ for $t_{\rm c}=0.05$, $0.2$, $0.5$ and $1\,H^{-1}$. The instantaneous-conversion times above are therefore representative. For $t_{\rm c}=H^{-1}$ the field never reverses, and the flux $b^2E$ has fallen to about $0.27$ of its initial value at the stopping surface $a=10^{-6}$ (to $0.271$ at $10^{-8}$ and $0.267$ at $10^{-10}$: unlike the endpoint time, the residual flux still drifts with the cutoff and is not a limit). The local field, which grows as $b^{-2}$ in the contracting throat, first falls to $0.67\,\EN$ and then grows far above its initial value. The removal of the charge, the local field and the field energy are distinct quantities, and it is the first of them that the criterion~\eqref{eq:criterion} refers to. For slower discharges the homogeneous equations still produce a crunch, $1.2\,H^{-1}$ for $t_{\rm c}=2H^{-1}$ and $1.8\,H^{-1}$ for $10H^{-1}$, because the Nariai throat is unstable to any homogeneous perturbation. These are the two runs whose step size underflows just before the stopping surface. In that regime, however, the inhomogeneous modes can no longer be neglected: they grow on a comparable time scale, up to a logarithm of their initial amplitude, and they can fragment the throat into black holes in de Sitter space instead of producing a global crunch~\cite{Bousso:1996au,Bousso:1998bn}. We adopt the criterion~\eqref{eq:criterion} as the condition that the conversion is complete before this happens. The logarithmic sensitivity of the homogeneous crunch time to $t_{\rm c}$ is one of the reasons for the $\pm3$ of Section~\ref{sec:TFLbound}. The static-patch description, in which the carriers are said to escape through the horizons, describes the same homogeneous process from the point of view of one observer, who sees the carriers of that patch leave and those of the neighboring patches arrive.

\paragraph{Pair creation at zero temperature.} Pairs are created at the rate per unit volume $\Gamma=(gq\EN)^2e^{-S}/4\pi^3$ (spin $\tfrac12$, and half of this for a complex scalar) and accelerated in opposite directions. After $t_{\rm acc}=m/(gq\EN)$ they move at essentially the speed of light and after a time of order $H^{-1}$ they cross one of the two horizons. The worldline instanton has radius $R=m/(gq\EN)=t_{\rm acc}$, so the flat-space treatment requires $RH<1$. At the bound~\eqref{eq:FL_log} this holds whenever $\lS>0$. The steady-state charge density of each sign is therefore $\Gamma/H$ up to an order-one factor, and the current density is $J_{\rm p}\simeq2gq\Gamma/H$. Setting $t_{\rm p}=\EN/J_{\rm p}<H^{-1}$ gives eq.~\eqref{eq:pair_criterion}. In the homogeneous description of the throat the carriers do not leave but accumulate. The accumulated-pair estimate of Section~\ref{sec:pair} agrees with the steady-state one up to $\ln2$ in $\lS$, and the quenching of the rate by the falling field brings the total shift to $\ln(\lS/2)$. Had we used only the $s$-wave current of the $dS_2$ analysis of~\cite{Frob:2014zka}, the pair-creation current would be smaller by the factor $\sim gq\MP/H$ that counts the transverse modes on the $S^2$. This would lower $\lS$ by $\ln(gq\MP/H)$ and make it negative for every $gq\lesssim1$ at any $H$, so that no zero-temperature bound would follow. Our four-dimensional estimate includes those modes. The residual uncertainty in the residence time shifts $\lS$ by an additive constant of order unity, which is irrelevant compared with $\ln(\MP/H)\simeq139$ at the present epoch.

\paragraph{Conduction.} The thermal population of the charged species, eq.~\eqref{eq:nth}, is accelerated by the field. Three conditions must hold for the ballistic estimate $J_{\rm c}=2gq\,n$ of Section~\ref{sec:conduction}. (i) The acceleration time must be shorter than the Hubble time, $t_{\rm acc}=m/(gq\EN)<H^{-1}$, i.e.\ $m<\sqrt6\,gq\MP$. For $gq=10^{-24}$ this is $m<6$\,keV. At the central normalization it holds down to $(gq)_{\min}$, and it fails only at the upper edge of the sensitivity band, where the current is multiplied by $e^3$, for couplings near and below $(gq)_{\min}$. The numerical results of Table~\ref{tab:mmin} and Figures~\ref{fig:massbound}, \ref{fig:neutrino_gBL} and \ref{fig:neutrino_charge} use the reduced mean velocity $\bar v=\min[1,\,gq\EN/(\max(m,3T)H)]$ throughout. (ii) Collisions with the bath must not prevent the particles from reaching relativistic velocities. For couplings of order $e$ the dominant process is Thomson scattering on the bath photons, with rate $\Gamma_{\rm sc}=n_\gamma\,8\pi\alpha^2/3m^2\simeq4\times10^{-11}$\,eV for $m=0.2$\,eV and $T=2.5$\,meV, i.e.\ a mean free time $\tau\simeq1.6\times10^{-5}$\,s, to be compared with $t_{\rm acc}\simeq6\times10^{-11}$\,s: the particles are re-accelerated to the speed of light between collisions and $\bar v=\mathcal O(1)$. For smaller couplings $\tau$ grows as $(gq)^{-4}$ while $t_{\rm acc}$ grows only as $(gq)^{-1}$. Even if the drift were collision limited, the current would only be reduced to the fraction $\tau/t_{\rm acc}$ of its ballistic value, which enters $\lc$ through its logarithm. (iii) The population must be in equilibrium, which is part of the definition of the states to which the conjecture applies. Under these conditions the field decays as $\dot E=-J_{\rm c}$ and $t_{\rm c}=\EN/J_{\rm c}$, eq.~\eqref{eq:cond_criterion}. Equivalently, since the particles cross the region between the horizons in a time of order $H^{-1}$, the criterion states that the plasma between the two horizons must contain at least $\sim\QN$ charge carriers of each sign, i.e.\ enough charge to neutralize the black hole in a single pass. The relevant screening length is the $\delta_{\rm s}=\EN/(gq\,n)$ of Section~\ref{sec:conduction}, not the equilibrium Debye length, which measures the linear response of the plasma to a small test charge. In the relativistic limit $m\lesssim T$ the density is $n=\tfrac34\tfrac{\zeta(3)}{\pi^2}g_sT^3$ for fermions ($\tfrac{\zeta(3)}{\pi^2}g_sT^3$ for bosons), which gives eq.~\eqref{eq:gqmin}.

\paragraph{Binary collisions and radiative losses.} Consider the carriers at the threshold density $n_{\rm th}=\sqrt6\MP H^2/(2gq)$, accelerated along the field to $\gamma(t)=t/t_{\rm acc}$ up to $\gamma_{\max}=gq\EN/(mH)$ ($10^{28}$ for the electron charge and $m=0.18$\,eV, $10^{5}$ for $g_{B-L}=10^{-24}$ and $m=43$\,meV). For two counter-streaming carriers the squared invariant mass is $s\simeq4\gamma^2m^2$, and the annihilation cross section, $\sigma\simeq(2\pi\alpha_Q^2/s)[\ln(s/m^2)-1]$, and the large-angle scattering cross section, $\sigma\sim4\pi\alpha_Q^2\ln\Lambda_{\rm C}/s$, are bounded by their values $\sim\sigma_T=8\pi\alpha_Q^2/3m^2$ near threshold and fall as $1/\gamma^2$ once $\gamma\gg1$ (bremsstrahlung is treated separately below). The optical depth accumulated along the trajectory is therefore dominated by the acceleration phase, $\tau_{\rm bin}=\int n\sigma\,\dd t\simeq n_{\rm th}\sigma_Tt_{\rm acc}$ up to a logarithm, which is eq.~\eqref{eq:taubin}: $\tau_{\rm bin}=(gq)^2H/(12\pi m)=1.6\times10^{-35}$ for $gq=e$, $m=0.18$\,eV, and $7\times10^{-82}$ for $g_{B-L}=10^{-24}$, $m=43$\,meV. Even the most generous estimate, $\sigma_T$ during the whole Hubble time, gives $\tau=n_{\rm th}\sigma_T/H=(gq)^3\EN/(12\pi m^2)$, i.e.\ $1.6\times10^{-7}$ for the electron charge and $10^{-76}$ for $g_{B-L}=10^{-24}$. Bremsstrahlung on the other carriers, whose cross section $\sigma_{\rm rad}\simeq(16/3)\alpha_Q^3\ln(\ldots)/m^2$ does not fall with $\gamma$, removes over a Hubble time the fraction $n_{\rm th}\sigma_{\rm rad}/H\simeq\tfrac83\alpha_Q^3\EN\ln(\ldots)/(gq\,m^2)$ of the energy, i.e.\ $7\times10^{-10}$ times a logarithm of order ten for the electron charge (the logarithm depends on the screening of the carriers and on the energy-weighted emission spectrum, which we do not evaluate) and $(gq)^5$ times smaller for smaller couplings. Radiative losses on the bath are small as well. In the Thomson regime, $\gamma<m/T$, the energy radiated is a fraction $4\times10^{-4}$ of the energy gained for the electron charge. In the Klein--Nishina regime, $\gamma T/m\gg1$, the loss rate on a blackbody at temperature $T$ saturates at $P\simeq\tfrac{1}{16}\sigma_Tm^2T^2[\ln(4\gamma T/m)-\tfrac{11}{6}-0.15]$~\cite{Blumenthal:1970gc}, where $-0.15$ is the average of $\ln(\omega/T)$ over the photons. The fraction of the energy gained from the field that is radiated during the first Hubble time, with $\gamma=t/t_{\rm acc}$, is then
\begin{equation}\label{eq:fKN}
    f_{\rm KN}=\frac{\langle P\rangle}{gq\EN}=\frac{\pi}{6}\,\frac{\alpha_Q^2T^2}{gq\EN}\left[\ln\frac{4\gamma_{\max}T}{m}-3.0\right]=5\times10^{-3}\qquad(gq=e,\ T=\Tmax)\,,
\end{equation}
and smaller by $(gq)^3$ for smaller charges (for $gq\lesssim10^{-29}$ the Klein--Nishina regime is not even reached and the losses are Thomson-like and utterly negligible). Neither collisions nor radiation affect the discharge at threshold. A collisional thermalization of the carriers, if it were required, would need the density $n_{\rm th}/\tau_{\rm bin}$ discussed in Section~\ref{sec:TFLbound}.

\paragraph{Collective dynamics.} The carriers are dilute, but their dynamics is collective. At the threshold density the plasma frequency of the two species is $\omega_p^2=2(gq)^2n_{\rm th}/m=\sqrt6\,gq\MP H^2/m=H/t_{\rm acc}$, so that $\omega_p/H=(t_{\rm acc}H)^{-1/2}$, which is $10^{14}$ for the electron charge, $4\times10^{2}$ for $g_{B-L}=10^{-24}$ and still $2.4$ for $g_{B-L}=10^{-29}$. Because the carriers become relativistic within $t_{\rm acc}$, the discharge is not a small-amplitude plasma oscillation but the relativistic one of Section~\ref{sec:conduction}: the field decreases linearly, $E(t)=\EN-J_{\rm c}t$, vanishes at $t_{\rm c}$ and reverses, the carriers turn around at $2t_{\rm c}$, and the period is $4t_{\rm c}$. Over a period the pressure along the field lines averages to $\langle p_\parallel\rangle=+\rho_E(0)/3$, which is what enters eq.~\eqref{eq:KS_collapse}. These are frozen-geometry expressions, valid while the throat has not yet reacted. In the self-consistent evolution above, the geometry changes both the reversal times and their number. The first field reversal is slightly delayed, and the first reversal of the current, which follows the carrier momentum and not the field, slightly advanced: $0.0501$ and $0.0997\,H^{-1}$ against $0.05$ and $0.1$ for $t_{\rm c}=0.05H^{-1}$. The number of reversals before the crunch falls from eight at $t_{\rm c}=0.05H^{-1}$ to two at $0.2$, one at $0.5$ and none at $t_{\rm c}=H^{-1}$. The averaged pressure $\langle p_\parallel\rangle=\rho_E(0)/3$ therefore applies only when the oscillation is fast compared with the evolution of the geometry, which is no longer the case at threshold. There the coupled solution above, which does not use the average, is what supports eq.~\eqref{eq:KS_collapse}. The two cold counter-streaming beams are unstable to the electromagnetic filamentation (Weibel) instability~\cite{Weibel:1959zz,Medvedev:1999tu,Bret:2010ifn}, whose growth rate for symmetric relativistic beams is $\Gamma_W\simeq\beta\,\omega_p/\sqrt\gamma$. With $\gamma=t/t_{\rm acc}$ this gives $\Gamma_W^2\simeq2gq\,n/(\EN t)$, i.e.\ $\Gamma_Wt_{\rm c}=1$ at the discharge time for any density, and the number of $e$-folds of growth during the first discharge is $\int_{t_{\rm acc}}^{t_{\rm c}}\Gamma_W\dd t=\int_{t_{\rm acc}}^{t_{\rm c}}\dd t/\sqrt{t_{\rm c}t}=2\bigl(1-\sqrt{t_{\rm acc}/t_{\rm c}}\bigr)\leq2$ for any density (with the linear decrease of the field included in the acceleration, $\pi/\sqrt2\simeq2.2$). The unstable wavelengths, of order $\Gamma_W^{-1}\lesssim t_{\rm c}$, fit inside the throat. Two $e$-folds amplify the seed fluctuations by less than an order of magnitude. The estimate $\gamma=t/t_{\rm acc}$ does not hold beyond the first reversal of the field, and the growth during the subsequent oscillations would require a calculation on the oscillating background with a seed and a saturation criterion (see e.g.~\cite{Bret:2013qva} for the fluctuation-to-saturation treatment in the collisionless-shock context). The beams may therefore isotropize collectively, but our estimate does not establish the ``radiation cloud'' of~\cite{Montero:2019ekk}. Since the collapse of the sphere does not depend on this isotropization, eq.~\eqref{eq:KS_collapse}, the discharge criterion is unaffected in either case.

\paragraph{Thermal pair creation.} For $T\gg\Tc$ the lens instanton describes a pair created from two thermal photons and pulled apart by the field~\cite{Brown:2015kgj,Gould:2017fve}. The perturbative rate of that process, with its prefactor, was computed in~\cite{Torgrimsson:2019sjn}, whereas~\cite{Gould:2018ovk} treats the worldline-sphaleron regime. In the weak-field limit the rate follows from detailed balance: in equilibrium the creation of pairs from the photons balances their annihilation, so that $\Gamma_T=n_+n_-\langle\sigma v\rangle$ with $n_\pm=g_s(mT/2\pi)^{3/2}e^{-m/T}$ and, for a spin-$\tfrac12$ pair at rest, $\langle\sigma v\rangle=\pi\alpha_Q^2/m^2$, i.e.\ $\Gamma_T=\alpha_Q^2mT^3e^{-2m/T}/2\pi^2$. The field enters only through the exponent $S_T$ of eq.~\eqref{eq:HighT_correct}, and this is the weak-field limit of the rate of~\cite{Torgrimsson:2019sjn}. The naive counting $\sigma_{\gamma\gamma}\sim\alpha_Q^2/m^2$ times the density $\sim m^2Te^{-m/T}$ of photons above threshold overestimates the rate by the factor $2\pi^2m/T\simeq10^{3}$, because the fusion cross section vanishes at threshold, in proportion to the velocity of the pair. Since the argument of the logarithm scales as $(gq)^5$, the solution drops below $m\simeq10\,T$ for $gq\lesssim10^{-10}$ at the present epoch. Below that value the non-relativistic rate is not applicable, and the disappearance of its formal roots at $gq\simeq5\times10^{-12}$ (where $2x=\ln(Dx)$ with $x=m/T$ ceases to have a solution, at $x=1/2$) carries no physical information.

\paragraph{Comparison of the channels.} Subtracting the high-temperature form of eq.~\eqref{eq:pair_thermal}, $m/T\geq\lS^{(T)}/2$, from eq.~\eqref{eq:TFL} gives eq.~\eqref{eq:compare}, $\lc-\lS^{(T)}/2=\ln[2\sqrt{2\pi}\,g_sm/((gq)^{3/2}\sqrt{\EN})]$ (for $\bar v=1$), e.g.\ $8.3$ for $gq=e$ and $m=0.18$\,eV, and much larger for smaller couplings. The comparison holds in the common domain of validity of the two expressions, $m\geq10T$. Below it the non-relativistic photon-fusion rate is outside its range and we draw no conclusion from it, and other thermal channels are not evaluated here.

\section{SMEFT formulae}\label{app:smeft}
With the operators of eq.~\eqref{eq:dim_6_SMEFT} and $H=(0,h/\sqrt2)^T$, the tree-level Lagrangian of the radial mode is $\mathcal L=\tfrac12Z(h)(\partial h)^2-V(h)$ with $Z(h)=1-2C_{H\Box}h^2/\Lambda^2+C_{HD}h^2/2\Lambda^2$ and $V(h)$ as in eq.~\eqref{eq:V_CH}. The vacuum expectation value $v$ is defined by $V'(v)=0$ and is the parameter that enters the gauge-boson masses, $m_W=gv/2$. The physical Higgs mass is the mass of the canonically normalized fluctuation, $m_h^2=V''(v)/Z(v)$. Solving these two conditions for $\mu^2$ and $\lambda$ gives eq.~\eqref{eq:smeft_params}, and $V(0)$ follows from $V(v)=0$. The one-loop thermal potential contains $\tfrac12\alpha T^2h^2$, with the thermal-mass coefficient $\alpha=\tfrac{1}{16}(3g^2+g'^2)+\tfrac{\lambda}{2}+\tfrac{y_t^2}{4}$~\cite{Carrington:1991hz,Arnold:1992rz}, where the scalar contribution follows from the field-dependent Higgs and Goldstone masses, $m_h^2(h)+3m_G^2(h)=-4\mu^2+6\lambda h^2+\mathcal O(h^4)$, and the top contribution from $m_t^2(h)=\tfrac12h^2(y_t-C^{33}_{uH}h^2/2\Lambda^2)^2$, whose $h^2$ coefficient is $y_t^2/2$. With $m_t=\tfrac{v}{\sqrt2}(y_t-C^{33}_{uH}v^2/2\Lambda^2)$ fixed to its measured value, $y_t$ is given by eq.~\eqref{eq:yt_smeft}. Expressing $g$ and $g'$ through $m_W$ and $m_Z$ (the $\mathcal O(C_{HD}v^2/\Lambda^2)$ shift of $m_Z$ is neglected, as is the $\mathcal O(Cv^2/\Lambda^2)$ modification of the scalar-loop term $\lambda/2$ through the rescaled propagators, which is a correction to the smallest term in $\alpha$) yields the bracket in eq.~\eqref{eq:smeft_master}. The symmetry-restoration temperature is $T_0^2=\mu^2/\alpha$, and the requirement $\sigma T_0^4\geq cV(0)$ is eq.~\eqref{eq:smeft_master}. For the Standard Model, $\alpha_{\rm SM}=0.398$, $T_0=140$\,GeV and $V(0)^{1/4}=104$\,GeV with $m_W=80.37$, $m_Z=91.19$, $m_h=125.2$, $m_t=172.6$ and $v=246.2$\,GeV. The conversion of the Higgs-pair result of~\cite{CMS:2026nuu} into a bound on $C_H$ uses $\kappa_\lambda=V'''(v)/(3m_h^2/v)=1-2C_Hv^4/(m_h^2\Lambda^2)$. The thermal mass used here is a selected leading-order approximation: it keeps the exact tree-level dependence on $C_H$ but omits the corrections listed above, which are of the same order in the effective-theory expansion, and it uses the high-temperature expansion without daisy resummation. The diagnostic boundary is accurate only to that order, which is one of the reasons we do not read its failure as an exclusion.

\section{Reheating}\label{sec:reheatingtop}
\subsection{The reheating bath and the electron}\label{sec:reheating}

A thermal bath with $\rho_R\simeq\rho_V$ is realized at the end of inflation, when the inflaton energy is converted into radiation. For instantaneous reheating, eq.~\eqref{eq:Inst_reh}, the radiation energy after the conversion equals the inflaton energy before it, i.e.\ $T_{\rm reh}=\Tmax(V_I)$ with $c=1$. The two are never present at the same time, which is already a sign that the conjecture does not apply here. We nonetheless follow the formal argument through. If the thermal FL conjecture of Section~\ref{sec:criterion} applied to this state, every particle charged under electromagnetism would have to satisfy eq.~\eqref{eq:TFL_final} with $T=T_{\rm reh}$ and $H=H_I$. For a particle of charge $e$, with $H_I^2=\sigma_{\rm reh}T_{\rm reh}^4/3\MP^2$, the logarithm is
\begin{equation}\label{eq:ell_reh}
    \lc=\ln\!\left[\frac{12\,e\,(mT_{\rm reh}/2\pi)^{3/2}\MP}{\sqrt6\,\sigma_{\rm reh}\,T_{\rm reh}^4}\right]\simeq49\text{--}57\qquad(T_{\rm reh}=10\,{\rm MeV}\text{--}10\,{\rm keV}),
\end{equation}
where the numerical values are those at the solution $m=\lc T_{\rm reh}$, i.e.\ the smallest allowed mass is $m_{\min}\simeq50\,T_{\rm reh}$. The electron satisfies $m_e\geq\lc(m_e,T_{\rm reh})\,T_{\rm reh}$ only for
\begin{equation}\label{eq:Treh_bound}
    T_{\rm reh}\;\lesssim\;9\ {\rm keV}\,,
\end{equation}
where we used $g_*=3.36$ ($10.75$ above $e^+e^-$ annihilation). For $T_{\rm reh}=4$\,MeV, close to the lowest value compatible with nucleosynthesis ($T_{\rm reh}\gtrsim2$--$5$\,MeV from the light-element abundances alone, depending on the decay channel of the reheating particle, and $T_{\rm reh}\gtrsim6$\,MeV once CMB and large-scale-structure data are included, at 95\% CL~\cite{Hasegawa:2019jsa,Barbieri:2025moq}), the electron would have to be heavier than $0.2$\,GeV, and above the electroweak crossover, $T_{\rm reh}>159$\,GeV, its Higgs-generated mass vanishes altogether (the thermal self-energy $\sim eT$ is far below $\lc T$). At $0.2$\,GeV the electron leaves no $e^\pm$ pairs in an MeV bath, and a Higgs displaced far enough to give it that mass raises the masses of all the other quarks, leptons and weak bosons by the same factor of about $400$: the displacement that would rescue the bound spoils the standard nucleosynthesis it was invoked to protect. The electron is not special: at $T_{\rm reh}=4$\,MeV the muon fails as well, and since $m_{\min}\simeq40$--$50\,T_{\rm reh}$ every charged fermion of the Standard Model and the $W$ boson fail for $T_{\rm reh}\gtrsim5$\,GeV. The ratio $m_{\min}/T$ quoted here is the one at MeV--GeV temperatures and is not universal. At the highest inflationary scales the same formal expression gives $m_{\min}/T$ of order unity ($1.8$ at $H_I=4.7\times10^{13}$\,GeV), and above $T\simeq\sqrt6\,eF(0)\MP/\sigma=9\times10^{15}$\,GeV, where $F(0)=3\zeta(3)/2\pi^2$ is the massless density coefficient and $H\simeq1.2\times10^{14}$\,GeV, not even a massless species saturating the current can discharge the black hole. There the criterion is empty rather than violated. Put differently, at MeV--GeV temperatures the formal application of the criterion to the reheating plasma, evaluated with $H=H_I$, is violated by the Standard Model spectrum by many orders of magnitude, which is the reason it cannot be imposed on reheating without a separate dynamical derivation.

A Higgs displacement persisting into reheating would formally remove the conflict ($\phi_{\rm reh}\gtrsim2.4\times10^{4}\,{\rm GeV}\times T_{\rm reh}/{\rm MeV}$ in eq.~\eqref{eq:ell_reh}), but it is not available. The displacement needed to give the electron the mass $m_e=xT_{\rm reh}$, $\phi=\sqrt2\,xT_{\rm reh}/y_e$, stores the energy $\lambda\phi^4/4$, and requiring it to stay below the energy of the plasma itself, $\rho_R=\sigma T_{\rm reh}^4$, gives $\lambda\leq\sigma y_e^4/x^4$: at $T_{\rm reh}=4$\,MeV, where $x=50$ and $\phi=9.6\times10^{4}$\,GeV, this is $\lambda\leq4\times10^{-29}$, whereas the Standard Model quartic would store $10^{27}$ times the energy of the plasma. Independently of the energy, a displacement of the Standard Model Higgs is also short lived: with $m_h/H\simeq2\times10^{25}$ at $T_{\rm reh}=4$\,MeV small oscillations have the period $2\pi/m_h\simeq3\times10^{-26}$\,s and the lifetime $\Gamma_h^{-1}\simeq2\times10^{-22}$\,s, shorter than $H^{-1}$ by twenty orders of magnitude (for large amplitudes the frequency and the decay rate are field dependent and the plasma contributes, which we do not model). Finally, a Higgs held at $10^{5}$\,GeV would give every charged Standard Model particle a mass far above $T_{\rm reh}$, leaving a plasma of photons and neutrinos (and the gluon sector in its confined form) only. As we now argue, the rescue is also not needed.

\subsection{Domain of validity of the conjecture}\label{sec:validity}

Taking the Standard Model with $T_{\rm reh}\gtrsim2$--$6$\,MeV~\cite{Hasegawa:2019jsa,Barbieri:2025moq} to be a consistent low-energy theory, the conflict just described shows that the thermal FL conjecture cannot be applied to the reheating bath. The physical reason is clear: the FL argument needs a vacuum-dominated background that lasts for the evolution of the black hole, so that the Nariai solutions exist and the conversion of electric-field energy leads to a crunch of the $S^2$ of the Nariai throat~\cite{Montero:2019ekk} (Section~\ref{sec:criterion}). During reheating the ``vacuum energy'' is the energy of the inflaton, which is itself being converted into the bath within a Hubble time. There is no static charged Nariai solution to speak of, and the fate of a large charged black hole formed at the end of inflation is that of a black hole in a radiation-dominated Friedmann universe. We therefore restrict the conjecture to backgrounds that satisfy two separate requirements: the vacuum energy must change negligibly over the time scale of the black-hole evolution, $|\dot V|/(HV)\ll1$, and the near-ultracold configuration must exist for that time, which requires the vacuum energy to dominate. Slow roll, with $\epsilon_H\ll1$, satisfies both. Adiabaticity of $V$ alone is not sufficient, since a constant $V$ accompanied by $\rho_R=V$ already gives $\epsilon_H=1$. During reheating neither requirement is met: the inflaton energy is being converted, and $\epsilon_H$ is of order unity and model dependent (for an inflaton oscillating in $V\propto\phi^n$ the virial theorem gives $\langle w\rangle=(n-2)/(n+2)$~\cite{Turner:1983he}, so that the time average of $\epsilon_H=\tfrac32(1+w)$ is $3n/(n+2)$, i.e.\ $1.5$ for $n=2$ and $2$ for $n=4$, and a radiation-dominated era approaches $2$). The domain so defined includes the present vacuum energy, to which Section~\ref{sec:particlephysics} applies, and slow-roll inflation, to which Section~\ref{sec:slowroll} applies, but excludes the reheating transition, which is non-adiabatic and outside the present calculation. A bath at the formal benchmark $c=1$ does not meet the second requirement either, since the bath alone collapses the throat within $0.77/H$ (Section~\ref{sec:criterion}), while for $c\lesssim0.5$ the throat survives a Hubble time. This is why the $c=1$ numbers of Section~\ref{sec:particlephysics} are a formal envelope, and why their dependence on $c$ is given there. The same restriction removes any application of the thermal bound to the reheating plasma, as well as any bound on the reheating temperature or on the inflationary scale that one might have derived from eqs.~\eqref{eq:ell_reh} and~\eqref{eq:Treh_bound}: within the domain of the conjecture, the constraints on inflation are those of Section~\ref{sec:slowroll}. The zero-temperature bounds of~\cite{Abu-Ajamieh:2025sul}, which apply the FL bound during inflation at the lowest inflationary scale allowed by nucleosynthesis, are statements about the inflationary epoch and are not affected by this restriction. They hold with the logarithm of eq.~\eqref{eq:FL_log} and under the condition stated in Section~\ref{sec:mCP}.

\section{Higgs inflation}\label{sec:higgsreview}
In the following, we give a brief review of the $\lambda \phi^4$-inflation with non-minimal gravitational coupling. 
Suppose a Higgs field $\Phi$ in a certain gauge theory plays the role of the inflaton. As a simple example, we may consider the Abelian Higgs model. An extension of the following discussion to the Standard Model with the Higgs doublet $\Phi$ is straightforward.  

The action of the inflaton field is defined in the so-called Jordan frame as follows 
 (we work in Planck units, namely, the reduced Planck mass $M_P=2.4 \times 10^{18}$\,GeV is set equal to 1):  
\begin{eqnarray}
 {\cal S}_J &=& \int d^4 x \sqrt{-g} 
   \left[-\frac{1}{2} f(\Phi^\dagger \Phi)  {\cal R}+  g^{\mu \nu} \left(\partial_\mu \Phi^\dagger  \right) \left(\partial_\nu \Phi \right)  - V_J  \right]  , 
\label{S_J}
\end{eqnarray}
where $f(\Phi^\dagger \Phi) = 1+ 2 \,  \xi \, (\Phi^\dagger \Phi )$ with a real parameter $\xi \geq 0$, 
  and the inflaton (Higgs) potential is given by
\begin{equation}   
V_J = \lambda \left(\Phi^\dagger \Phi - \frac{v^2}{2} \right)^2 .
\label{eq:infpot}
\end{equation}

Here, we have dropped the gauge interaction terms since they are irrelevant to the discussion about inflation. 
During inflation, the gauge symmetry is broken, and in the unitary gauge, we express the action of the inflaton as 
\begin{eqnarray}
 {\cal S}_J &=& \int d^4 x \sqrt{-g} 
   \left[-\frac{1}{2} f(\phi)  {\cal R}+ \frac{1}{2} g^{\mu \nu} \left(\partial_\mu \phi \right) \left(\partial_\nu \phi \right)  - V_J (\phi) \right]  , 
\end{eqnarray}
where $\phi=\sqrt{2} \Re[\Phi]$, $f(\phi) = 1+ \xi \, \phi^2$, and the inflaton potential is approximately given by  
$V_J \simeq \frac{1}{4}\lambda \phi^4$ since during inflation, $\phi \gg v$. 
For $\xi = 0$, this action reduces to the case of standard chaotic inflation with a quartic potential. 

The action in the Jordan frame can be transformed into the so-called Einstein frame with the conformal transformation $f(\phi) g_{\mu \nu} =  g_{E {\mu \nu}}$, 
\begin{eqnarray}
S_E &=& \int d^4 x \sqrt{-g_E}\left[-\frac{1}{2}  {\cal R}_E +  \frac{1}{2} g_E^{\mu \nu} \left(\partial_\mu \sigma \right) \left(\partial_\nu \sigma \right)-V_E(\phi(\sigma)) \right], 
\label{S_E}   
\end{eqnarray}
where the canonically normalized scalar $\sigma$ (inflaton in the Einstein frame) is related to the original field $\phi$ by  
\begin{eqnarray}
\left(\frac{d\sigma}{d\phi}\right)^{2} = \frac{1+ \xi (6 \xi +1) \phi^2} {\left( 1 + \xi \phi^2 \right)^2}, 
\label{eq:sigphi}
\end{eqnarray}
and 
\begin{eqnarray}
V_E(\phi(\sigma)) = \frac{V_J (\phi(\sigma))}{f(\phi(\sigma))^2} 
\simeq \frac{\frac{1}{4} \lambda \phi(\sigma)^4}{(1+\xi \, \phi(\sigma)^2)^2} \to \frac{\lambda}{4 \xi^2}  
\; \; {\rm for} \; \xi \, \phi(\sigma)^2 \gg 1.
\end{eqnarray}

\begin{table}[!t]
\begin{center}
\begin{tabular}{|c||ccc|c|c|}
\hline 
 $\xi $ & $n_s$   &  $r$  &  $\alpha (10^{-4})$  & $\lambda(\phi_I)$ & $\phi_I$\\ 
\hline
 $0$           &  $0.951$  &  $ 0.261 $      & $-8.06$      &  $ 1.38 \times 10^{-13}  $  & 22.1 \\
 $0.0134$  &  $0.965$  &  $ 0.0360 $      & $-6.02$      &  $ 1.09 \times 10^{-12}  $  & 21.5 \\
 $0.1$        &  $0.967$  &  $ 0.00784 $      & $-5.39$      &  $ 1.04 \times 10^{-11}  $  & 17.8 \\
$1$        &  $0.968$  &  $ 0.00346 $      & $-5.25$      &  $ 4.42 \times 10^{-10}  $  & 8.52 \\
$10$        &  $0.968$  &  $ 0.00301 $      & $-5.24$      &  $ 3.84 \times 10^{-8}  $  & 2.89 \\
$1000$        &  $0.968$  &  $ 0.00296 $      & $-5.23$      &  $ 3.77 \times 10^{-4}  $  & 0.291  \\
$10000$        &  $0.968$  &  $ 0.00296 $      & $-5.23$      &  $ 3.77 \times 10^{-2}  $  & 0.0921 \\
$25000$        &  $0.968$  &  $ 0.00296 $      & $-5.23$      &  $ 0.236  $  & 0.0583 \\
 \hline
\end{tabular}
\end{center}
\caption{ 
Inflationary predictions for various benchmark $\xi$ values and $N=60$. 
Values of $\xi < 0.0134$ ($r > 0.036$) are excluded by the BICEP/Keck results \cite{BICEPKeck:2022mhb}. } 
\label{Tab:2}
\end{table}

Note that $V_E$ approaches a flat potential as $\phi$ becomes larger ($\xi\phi^2>1$). 
This behavior is suitable for inflation. 
Using the relation between $\sigma$ and $\phi$ in eq.~\eqref{eq:sigphi}, 
  the inflationary slow-roll parameters in the Einstein frame are listed below:  
\begin{eqnarray}
 \epsilon(\phi) &=& \frac{1}{2} \left(\frac{V_E^\prime}{V_E \; \sigma^\prime}\right)^2,   \nonumber \\
 \eta(\phi) &=& \frac{V_E^{\prime \prime}}{V_E \; (\sigma^\prime)^2}- \frac{V_E^\prime \; \sigma^{\prime \prime}}{V_E \; (\sigma^\prime)^3} ,   \nonumber \\
 \zeta (\phi) &=&  \left(\frac{V_E^\prime}{V_E \; \sigma^\prime}\right) 
 \left( \frac{V_E'''}{V_E \; (\sigma^\prime)^3}
-3 \frac{V_E'' \; \sigma''}{V_E \; (\sigma^\prime)^4} 
+ 3 \frac{V_E^\prime \; (\sigma^{\prime \prime})^2}{V_E \; (\sigma^\prime)^5} 
- \frac{V_E^\prime \; \sigma'''}{V_E \; (\sigma')^4} \right),
\label{eq:SR1} 
\end{eqnarray}
where derivatives of $V$ with respect to $\phi$ are denoted by the {\it prime}.  
Inflation takes place if the slow-roll parameters satisfy $\epsilon,  |\eta|, |\zeta| \ll 1$. 

The amplitude of the curvature perturbation ($\Delta_{\cal R}^2 $) and the number of $e$-folds ($N$) are expressed as 
\begin{equation} 
  \Delta_{\cal R}^2 = \left. \frac{V_E (\phi)}{24 \pi^2 \epsilon (\phi) } \right|_{\phi_I},
\qquad 
N = \frac{1}{\sqrt{2}} \int_{\phi_E}^{\phi_I}
  d \phi  \frac{\sigma^\prime (\phi)}{\sqrt{\epsilon(\phi)}} \;,
\label{eq:SR2}
\end{equation}
where $\phi_I$ denotes the inflaton field value at the horizon exit scale, and $\epsilon(\phi_E)=1$ defines the end of inflation. 
From the Planck 2018 result \cite{Planck2018}, $\Delta_\mathcal{R}^2= 2.1 \times10^{-9}$ for the pivot scale $k_0 = 0.05$\,Mpc$^{-1}$, which corresponds to $\phi_I$. 
To solve the horizon and flatness problems the number of $e$-folds is typically set to be $N \gtrsim 50$.  

For a fixed value of $N$ and $\xi$, the three parameters, $\lambda$, $\phi_I$ and $\phi_E$, are determined 
  from eqs.~\eqref{eq:SR1} and~\eqref{eq:SR2} and $\epsilon(\phi_E)=1$
  with the Planck 2018 measurement, $\Delta_\mathcal{R}^2= 2.1 \times10^{-9}$. 
The inflationary predictions for the spectral index ($n_{s}$), tensor-to-scalar ratio ($r$), 
  and running of the spectral index ($\alpha=\frac{d n_{s}}{d \ln k}$) are given by 
\begin{eqnarray}
n_s = 1-6\epsilon+2\eta, \; \;
r = 16 \epsilon,  \;  \;
\alpha =16 \epsilon \eta - 24 \epsilon^2 - 2 \zeta, 
\end{eqnarray} 
where all three slow-roll parameters are evaluated at $\phi=\phi_I$. 

\begin{figure}[!t]
\begin{center}
\includegraphics[width=0.7\textwidth]{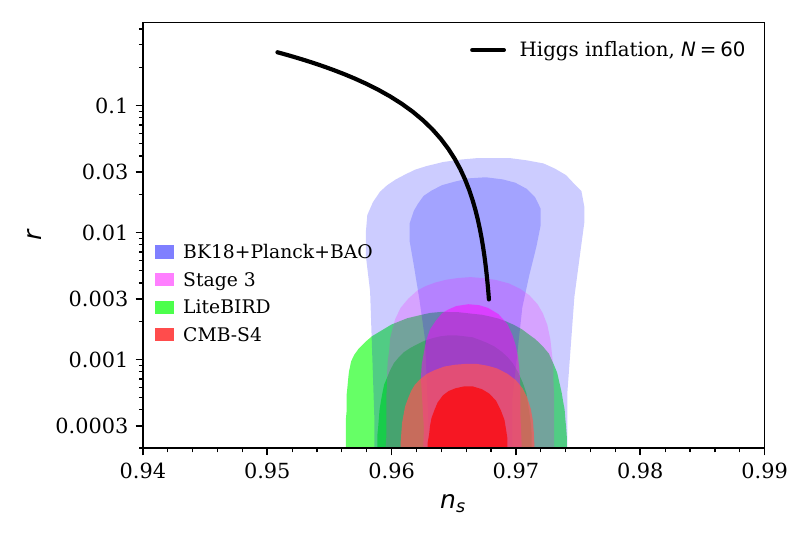} 
\end{center}
\caption{
The plot shows the inflationary predictions for $N=60$ along with the current best constraints from a combination of the BICEP2/Keck Array (BK18) experiments, Planck 2018 and BAO data \cite{BICEPKeck:2022mhb} and the search reach of future experiments such as LiteBIRD \cite{LiteBIRD:2022cnt} and CMB stage 3 and stage 4 (CMB-S4) \cite{Abazajian:2019eic}. 
}
\label{fig:infexp}
\end{figure}

The results for $N=60$ are summarized in Table~\ref{Tab:2}.
The values of $\xi < 0.0134$ ($r > 0.036$) are excluded by the BICEP/Keck results \cite{BICEPKeck:2022mhb}. 
The inflationary predictions for $n_s, r$, and $\alpha$ rapidly approach their asymptotic values for $\xi \gtrsim 1$, for example, see Figure~\ref{fig:infexp}, where we show the inflationary predictions for $N=60$ along with the current best constraints from the combination of the BICEP2/Keck Array 2018 experiments, Planck 2018 and BAO data \cite{BICEPKeck:2022mhb} and the search reach of future experiments such as LiteBIRD \cite{LiteBIRD:2022cnt} and CMB stage 3 and stage 4 (CMB-S4) \cite{Abazajian:2019eic}. 
For $\xi \gg 1$, $r$ approaches the constant value $r \simeq 12/N^2$ at leading order in $1/N$ for a fixed $N$ value, and $n_s$ can be approximated as 
\begin{eqnarray}
n_s \simeq 1-\sqrt{\frac{r}{3}}-\frac{3r}{8}. 
\end{eqnarray}

\end{document}